\documentclass[twocolumn]{aastex7}
\usepackage{amsmath}

\newcommand{\msun}{M$_\odot$}
\newcommand{\yt}{\citetalias{yt20}}
\newcommand{\dmunits}{~pc~cm$^{-3}$}
\newcommand{\ovii}{\ion{O}{7}}
\newcommand{\hi}{\ion{H}{1}}
\newcommand{\hii}{\ion{H}{2}}
\newcommand{\ha}{H$\alpha$}
\newcommand{\gitlink}{\url{https://github.com/slucchini/l26halodm}}

\defcitealias{yt20}{YT20}

\makeatletter % Workaround to only color the year for cite commands
\patchcmd{\NAT@citex}
{\@citea\NAT@hyper@{%
		\NAT@nmfmt{\NAT@nm}%
		\hyper@natlinkbreak{\NAT@aysep\NAT@spacechar}{\@citeb\@extra@b@citeb}%
		\NAT@date}}
{\@citea\NAT@nmfmt{\NAT@nm}%
	\NAT@aysep\NAT@spacechar\NAT@hyper@{\NAT@date}}{}{}

\patchcmd{\NAT@citex}
{\@citea\NAT@hyper@{%
		\NAT@nmfmt{\NAT@nm}%
		\hyper@natlinkbreak{\NAT@spacechar\NAT@@open\if*#1*\else#1\NAT@spacechar\fi}%
		{\@citeb\@extra@b@citeb}%
		\NAT@date}}
{\@citea\NAT@nmfmt{\NAT@nm}%
	\NAT@spacechar\NAT@@open\if*#1*\else#1\NAT@spacechar\fi\NAT@hyper@{\NAT@date}}
{}{}
\makeatother

\begin{document}
	
	% \title{Low Milky Way Halo Dispersion Measures in the ENGAWA Simulations}
	\title{A New Model for the Milky Way Halo Dispersion Measure\\with the ENGAWA Simulations: Low DMs and Large Anisotropy}
	
	\author[0000-0001-9982-0241]{Scott Lucchini}
	\affiliation{Center for Astrophysics $|$ Harvard \& Smithsonian, 60 Garden Street, Cambridge, MA 02138, USA}
	\email{scott.lucchini@cfa.harvard.edu}
	
	\author[0000-0002-7587-6352]{Liam Connor}
	\affiliation{Center for Astrophysics $|$ Harvard \& Smithsonian, 60 Garden Street, Cambridge, MA 02138, USA}
	\email{liam.connor@cfa.harvard.edu}
	
	\author[0009-0008-5043-6220]{Samuel McCarty}
	\affiliation{Center for Astrophysics $|$ Harvard \& Smithsonian, 60 Garden Street, Cambridge, MA 02138, USA}
	\email{samuel.mccarty@cfa.harvard.edu}
	
	\author[0000-0001-8235-2939]{Ralf M. Konietzka}
	\affiliation{Center for Astrophysics $|$ Harvard \& Smithsonian, 60 Garden Street, Cambridge, MA 02138, USA}
	\email{ralf.konietzka@cfa.harvard.edu}
	
	\correspondingauthor{Scott Lucchini}
	\email{scott.lucchini@cfa.harvard.edu}
	
	\begin{abstract}

		Fast radio burst (FRB) dispersion measures (DMs) are powerful tracers of the low density universe, but interpretation is made difficult by the integrated nature of DM. In this paper, we analyze the new ENGAWA cosmological zoom-in simulation suite to constrain the DM contribution from the Milky Way (MW) halo. The ENGAWA simulations consist of four Milky Way-like galaxies with enhanced resolution in the circumgalactic medium, making them an excellent tool to probe the properties of the gaseous halo. The median all-sky DM from the galactic halos in ENGAWA span 19--39\dmunits, varying even at fixed feedback strength, halo mass, and $f_\mathrm{CGM}$. These simulations show that, with enhanced circumgalactic resolution, the mean halo DM values are in general lower with more anisotropy across the sky than previous models, reaching $<10$\dmunits\ towards the poles. Furthermore, by varying the solar position within the simulated galaxies, we have obtained an estimate of the uncertainty in DM given our location in the MW. We provide a Python package with MW halo DM values from the simulation accessible via API functions of Galactic longitude and latitude. These new, simulation-based MW halo DM estimates will provide a critical baseline for interpretations of future FRB observations and our understanding of the global gas distribution in the Universe.
		
	\end{abstract}
	
	
	\section{Introduction}
	
	The circumgalactic medium (CGM) is a critical component of galaxy evolution. It is a massive reservoir of material that mediates all of a galaxy's accretion and outflows. A continuous supply of gas from the intergalactic medium can refuel the galaxy to sustain its star formation, while explosive feedback from star formation and active galactic nuclei can expel and heat the surrounding gas, inhibiting future star formation. The properties and contents of a galaxy's CGM directly impacts the efficiency and outcome of these processes.
	
	In the absence of astrophysical feedback, one naively expects that a galaxy should contain a total mass of baryons equal to the cosmic baryon fraction scaled by its halo mass, $M_b=(\Omega_b/\Omega_m)\times M_{200}\approx 0.16\times M_{200}$. However, it is unclear if this is the case. Even for our own Milky Way (MW), it has been a longstanding question if our Galaxy retains its cosmic share of baryons. With a halo mass of $M_{200}\sim1-1.5\times10^{12}$~\msun, the cosmic baryon fraction implies $\sim2\times10^{11}$\msun\ in baryonic mass, of which stars and the interstellar medium (ISM) account for only $\sim6\times10^{10}$\msun\ ($\sim$30\%; \citealt{bland-hawthorn16}).
	
	% It is a longstanding question if the Milky Way's halo retains its cosmic share of baryons. The cosmic baryon fraction of $\frac{\Omega_b}{\Omega_m} \approx 16\%$ implies $\sim$$2\times10^{11}\,M_\odot$ baryonic mass for a Milky Way halo mass of $M_{200}\sim1\textrm{--}1.5\times10^{12}\,M_\odot$, of which stars and the ISM account for only $\sim$$6\times10^{10}\,M_\odot$ \citep{bland-hawthorn16}. 
	The hot ($T\sim2\times10^{6}$\,K) phase of the Galactic halo is detected in \ion{O}{7} and \ion{O}{8} absorption toward background quasars \citep{gupta12,fang15} and in soft X-ray emission \citep{miller15}, most recently mapped over half the sky by eROSITA \citep{locatelli24}. Inferring total mass from Oxygen lines requires extrapolating over several orders of magnitude in abundance, under assumptions about metallicity and temperature. Ram-pressure stripping of dwarf satellites \citep{grcevich09,salem15,putman21} and the dispersion of Magellanic and globular cluster pulsars \citep{anderson10} provide independent, metallicity-free density constraints at tens of kpc. Total mass estimates still vary widely, from a few $\times10^{10}\,M_\odot$ for steep density profiles \citep{miller15,bregman18} to $\gtrsim10^{11}\,M_\odot$ for flatter or cored models, which would account for the full expected CGM baryon budget \citep{gupta12,faerman17}.

	The dispersion measure (DM) of fast radio bursts (FRBs) provides a column density measurement of free electrons between source and observer, offering a new probe of diffuse, ionized baryonic matter \citep{lorimer2007,macquart2020}. The application of FRB DMs to mapping the cosmic baryons is a growing field, a central goal of which is to better understand the nature of the CGM. 
	DM is often partitioned in the following way,
	\begin{equation}
		\rm DM_{obs} = DM_{MW, ISM} + DM_{MW,H} + DM_{cos} + DM_{host},
	\end{equation}
	where $\rm DM_{MW, ISM}$ is the Galactic ISM contribution, $\rm DM_{cos}$ is the cosmic DM that arises from traversing the IGM and intersecting halo gas, and $\rm DM_{host}$ corresponds to dispersion from the FRB host galaxy and its halo. 
	
\begin{figure*}[t]
		\centering
		\includegraphics[width=1.0\textwidth]{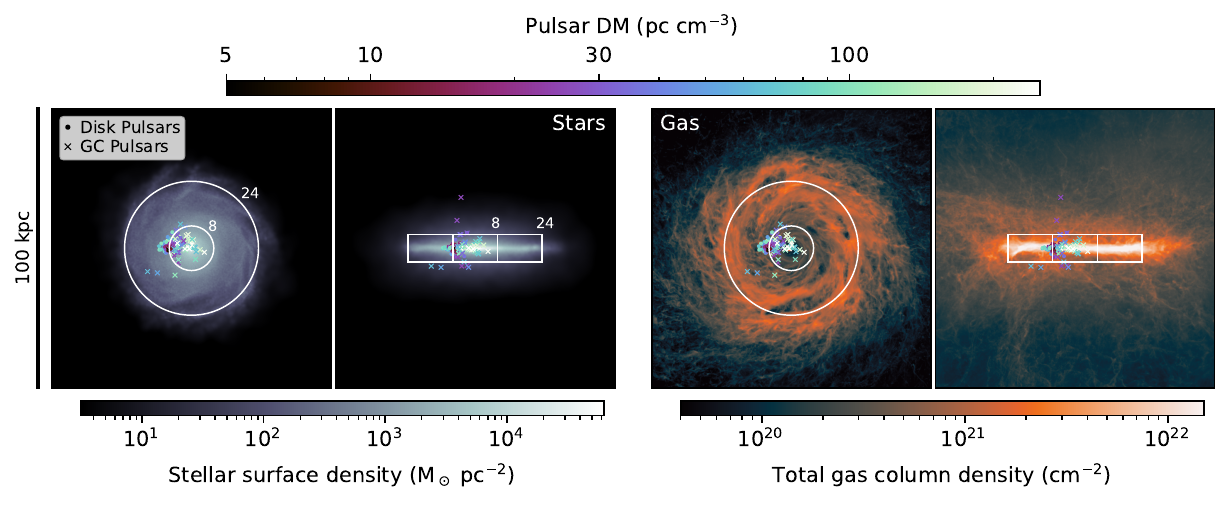}
		\caption{Face-on and edge-on projections of the stars (left) and gas (right) from the simulation (Au6 at 200~pc resolution), overlaid with the positions of the pulsars used to compute the local DM model for our galaxy which is excluded from our estimates. The circles and rectangles show radial extents of 8 and 24~kpc with heights of 10~kpc ($|z|<5$). For this simulation, all gas within the outer cylinder ($24\times10$~kpc) is not included in the halo DM calculation. This radius was calculated based on the edge-on density profile of the disk (see Section~\ref{sec:methods}).}
		\label{fig:pulsars}
	\end{figure*}

	Here, we are concerned with DM from the Milky Way's halo, $\rm DM_{MW,H}$. As FRB sample sizes grow, $\rm DM_{MW,H}$ will play an increasingly important role as a limiting systematic for FRB cosmology because it must be subtracted off of the total observed DM for inference. Beyond its role as a foreground, the baryon content of the MW is itself a key question in galaxy formation. The Galactic ISM DM is reasonably well modeled by pulsars, the spatial distribution of which is shown in Figure~\ref{fig:pulsars}. The halo DM and its angular distribution, on the other hand, remain poorly constrained. Understanding the PDF of DM$_{\rm MW,H}$ for the MW is critical both for FRB cosmology and understanding the CGM.
	
	Most FRBs discovered to date have come from beyond $\sim$\,Gpc where their DM is dominated by the IGM and intervening halo gas \citep{connor2025}. However, local universe FRBs are unlikely to intersect foreground galaxy halos, and receive little contribution from the IGM with DM$_{\mathrm{IGM}}\approx 10\,\left ( \frac{d_A}{50\,\rm Mpc} \right)\, \rm pc\,cm^{-3}$. Thus, FRBs at non-cosmological distances present a path towards measuring the integrated baryon content of the ionized CGM of the MW and of FRB host galaxies.
	
	Prior to having a direct probe of the total ionized gas content, metal tracers of the hot Galactic halo gas have been used as a proxy to estimate $\rm DM_{MW,H}$. In \citet{prochaskazhang}, the authors estimated $\rm DM_{MW,H}$ by extrapolating from measured \ion{O}{7} column densities as,
	\begin{equation}
		\mathrm{DM}_{\rm O_{VII}} \approx 80\,\rm pc\,cm^{-3}\left ( \frac{Z}{Z_\odot} \right )^{-1}\,\left ( \frac{N_{OVII}}{10^{17}\,cm^{-2}} \right ).
	\end{equation}
	
	The proposed range of 50--80\dmunits\ for $\rm DM_{MW,H}$ has been widely adopted in the FRB literature, as it was the first empirically motivated estimate. Later, \citet{yt20} (hereafter, \citetalias{yt20}) produced an analytic model for the halo DM, assuming a spherical component and a disk-like hot gas component to reproduce anisotropy in the observed X-ray emission measure (EM). Their average DM$_{\rm MW,H}$ is 48\dmunits\ (median 38\dmunits) with significantly higher values at low Galactic latitude. In \citet{keatingpen}, it was argued that X-ray observations were consistent with a wide range of MW halo DMs spanning an order of magnitude, including $\lesssim$\,10\dmunits. \cite{Das_2020} compiled 21-cm, UV, EUV and X-ray data to estimate DM$_{\rm MW,H}$, finding a median of 64\dmunits.
	
	Since then, nearby FRBs have been detected by multiple surveys and constraints on the baryon column through the Galactic halo have emerged using extragalactic DMs. \citet{ravi} found FRB\,20220319D at 50\,Mpc whose total DM was smaller than the ISM contribution in that direction as modeled by NE2001 \citep{ne2001} and YMW15 \citep{ymw15}. They obtained an upper-limit on the MW halo DM of 47.3\dmunits\ from this single sightline. Assuming spherical symmetry in the halo, this would imply that the MW's baryon budget is $\leq60\%$ of $(\Omega_b/\Omega_m)$. An analysis of unlocalized CHIME/FRB sources placed a limit of 52--111\dmunits\ \citep{Cook_2023}. More recently, \citet{mccarty2026} analyzed five local Universe FRBs that showed signs of low host-galaxy ISM contribution to DM, selecting for face-on galaxies, large offsets, low scattering measure, and low Faraday rotation measure.
	Their values for the combined DM of the host galaxy CGM and MW CGM fall below 40\dmunits\ for that sample, placing upper limits on the MW CGM DM contribution along those sightlines. \citet{hoffmann2026ihaloconstrainingmilky} used localized FRBs to determine $\rm DM_{MW,H}=68^{+27}_{-24}\,$\dmunits. Finally, \citet{liu2026investigatinganisotropydispersionmeasure} claim to have detected an anisotropy in halo DM by combining CHIME and localized FRBs. Taken together, there is not yet a consensus on the halo electron column from observations. 
	
	Compiling a cohesive model for the MW halo DM has proven difficult due to these observationally motivated models typically assuming a smooth, nearly-isotropic ionized gas distribution. This has led to a fairly uniform distribution of DMs, which is beginning to become inconsistent with the data. Thus, we have begun investigating hydrodynamic simulations to better understand the DM distribution in a realistic, cosmological setting. 
	% Furthermore, we note that observational efforts that use FRBs to constrain the baryon content of host galaxy halos and the Milky Way have typically used analytic models that assume a smooth, nearly-isotropic ionized gas distribution
	
	% Comparing these observational constraints with simulations is also 
	% In addition to the difficulties in obtaining direct constraints from observations, comparing with simulations is also 
	Unfortunately, deriving predictions from simulations is nontrivial due to the difficulty in simulating the large, diffuse component of galaxies. Traditionally, cosmological simulations have used a density-based criterion to focus computational energy on the galaxies themselves in order to overcome the vast hierarchy of scales required to simulate a cosmological region \citep{vogelsberger20}. This comes at the cost of low resolution in low-density regions, such as the CGM.
	
	Several additional techniques have been employed in modern zoom-in simulations to overcome this difficulty. By isolating the circumgalactic region using particle tracking or tracer fluids, additional refinement criteria can be activated around select galaxies without affecting the rest of the cosmological volume. This allows for much higher resolution simulations without a dramatic increase in computational cost. In the CGM region, one can increase the mass resolution \citep{suresh19,gible}, or the spatial resolution \citep{tempest,surge,foggie,engawa}; both techniques result in dramatic changes in the structure of the CGM. Higher resolution simulations find an increase in cool and cold material around galaxies \citep{surge,gible,engawa} with many more resolved cloudlets \citep{foggie10,engawa} and increased resolved turbulence and non-thermal support \citep{foggie5,foggie6}.
	
	\begin{figure*}[t]
		\centering
		\includegraphics[width=0.9\textwidth]{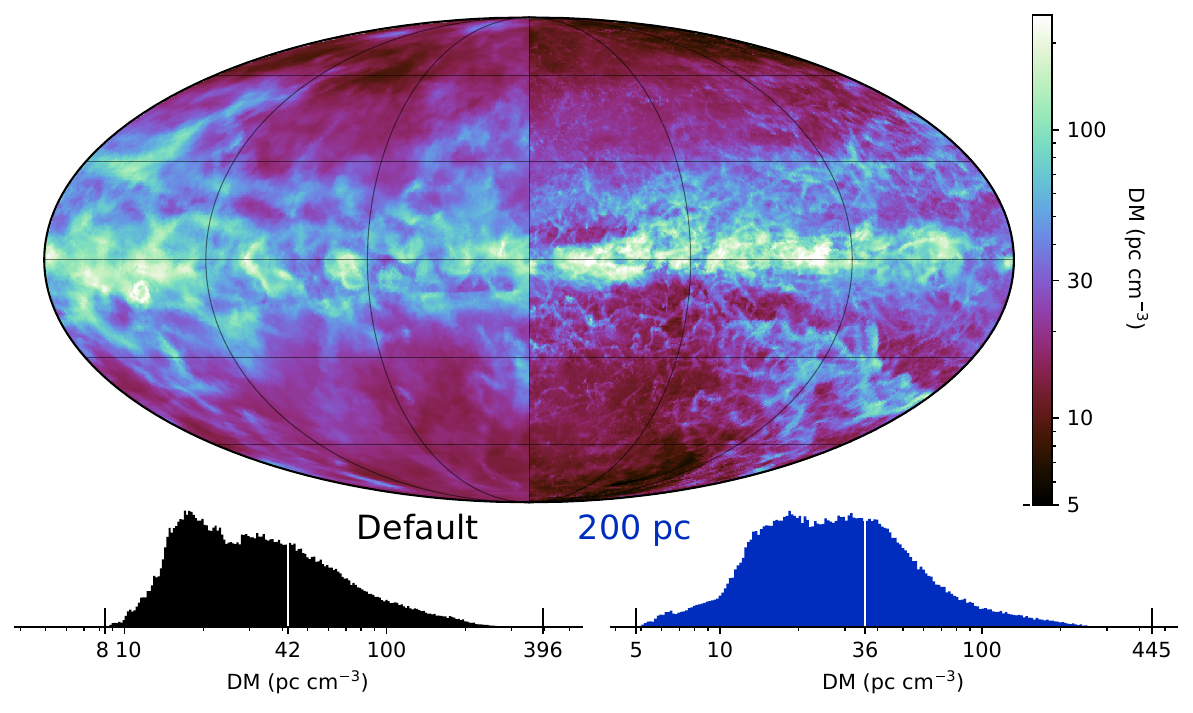}
		\caption{All-sky Mollweide projection of HEALPix DM maps for Au6 at Default and 200~pc resolution after masking the galactic disk. The bottom panels show the distribution of DMs weighted by sky area. The black vertical lines (and corresponding axis labels) denote the minimum and maximum DM value in the image, while the white lines (and labels) mark the means.}
		\label{fig:allsky}
	\end{figure*}
	
	Thus far, most efforts to simulate FRB DMs have focused on the cosmic web as well as the host halo DM contribution.
	For example, \citet{orr} used the Feedback in Realistic Environments (FIRE-2) cosmological zoom-in simulation suite to study DM from the host galaxy ISM for different stellar populations.
	In the context of the cosmic web, \cite{konietzka2025} recently presented a continuous ray-tracing framework to study FRB dispersion measures arising from the IGM and intervening halos in the IllustrisTNG simulation \citep{nelson19}. They found that both the simulation box size and resolution affects the resulting DM distribution of the cosmic web \citep{konietzka2025}. \citet{huang25} investigated the local universe ($<120$~Mpc) including the MW halo DM contribution, however their main focus remained on the local IGM contributions to the DM as well.
	% We will see later in this work that the MW halo contribution in simulations is also dependent on the underlying resolution.
	
	It is therefore timely to carry out high resolution hydrodynamical simulations of the CGM of MW-like galaxies. Cosmological zoom-in simulations allow us to characterize $\rm DM(\hat{\bf n})$ from the halo under different feedback assumptions, spatial and mass resolutions, and halo masses. Simulations also provide multiple tracers of the hot gas beyond dispersion, allowing for comparison against different probes.
	
	In this work we characterize the MW halo DM contribution with the ENGAWA simulation suite (ENhanced Galactic Atmospheres With Arepo; \citealt{engawa}). With fixed spatial resolution throughout the CGM, these simulations provide an excellent resource to probe the structure and contents of the CGM to better understand the results from these new probes. Section~\ref{sec:methods} describes the simulations and our analysis techniques, Section~\ref{sec:dms} shows the simulated all-sky DM results, Section~\ref{sec:obsv} compares to various observables to improve our DM predictive power, and in Section~\ref{sec:disc} we discuss these results and conclude.
	
	\section{The ENGAWA Simulations} \label{sec:methods}
	
	% \subsection{The ENGAWA Simulations}
	
	In this work we have used the high-resolution CGM-refinement ENGAWA simulation suite (ENhanced Galactic Atomspheres With Arepo; \citealt{engawa}). It consists of cosmological zoom-in simulations of four MW-like galaxies at four different resolutions with the Arepo multiphysics code \citep{springel10,arepo}. There are two galaxies selected from the Auriga suite \citep{grand17}, and two from the TNG50 simulation in the IllustrisTNG suite \citep{nelson19}. The galaxies have halo masses ranging from 1.0 to $1.4\times10^{12}$~\msun\ and stellar masses ranging from 3.0 to $9.2\times10^{10}$~\msun. The simulations employ the TNG galaxy formation model including the effective equation of state model for star formation, wind particle stellar feedback, and quasar-mode black hole feedback \citep{weinberger17,pillepich18}.
	
	All galaxies were simulated at Auriga Level 4 mass resolution by default ($m_\mathrm{baryon}=5.4\times10^4$~\msun). From redshift 0.3 to the present day (3.4~Gyr of evolution), an additional volume refinement criterion was added in the CGM specifying a maximum cell size. This was implemented at levels of 1~kpc, 500~pc, and 200~pc extending out to 100~kpc away from the galaxy (while maintaining the mass refinement if cells became dense enough). Note that there is no inner boundary for this refinement scheme $-$ it is active from $r=0$ out to $r=100$~kpc before smoothly relaxing back to full mass refinement at $r=200$~kpc. Thus, this suite is ideal for studying the inner CGM in high-resolution with the modern, well-tested TNG physics model.
	
	In \citet{engawa}, the $z=0$ snapshots were also post-processed with COLT (COsmic Lyman-alpha Transfer code; \citealt{colt}) which incorporates stellar radiation into the photoionization and collisional ionization equilibrium computation of the electron densities and ion fractions. Throughout this paper, we use the electron densities output from COLT instead of those directly from the simulation to account for the ionizing effect of stellar radiation. See \citet{engawa} for the specifics of the implementation.
	
	% \subsection{All Sky Ray-Tracing}
	
	To compute the MW halo DM in the simulations, we excised the galactic disks and integrated the electron density along a uniform HEALPix\footnote{\url{http://healpix.sourceforge.net}} distribution of lines of sight from a solar location out to the virial radius \citep{healpix,healpy}. To remove the disks, we calculated the edge-on total gas column density profile along the midplane of the galaxy ($|z|<5$~kpc), and determined the $x$ locations at which this profile dropped below $3.5\times10^{21}\ \mathrm{cm}^{-2}$. We used the average of the positive and negative $x$ locations as the radial limit of the disk. These values range from 19 to 32~kpc depending on the galaxy and the resolution. We used $|z|<5$~kpc as the vertical limit of the disk for all simulations.
	
	Figure~\ref{fig:pulsars} shows the distribution of stars and gas for one of the galaxies (Au6), compared with the locations of MW pulsars that have been used to constrain the DM of the ISM \citep{ne2001,ne2025}. The circles and rectangles show reference radii of 8 and 24~kpc (with heights of 10~kpc) with the outermost shape showing the disk cylinder cutout used for this galaxy. Most of the pulsars lie within this cylinder region and while some of the gaseous disk material appears outside this 25~kpc radius cylinder, viewing the galaxy edge on or viewing its stellar distribution, the dominant component of the disk is contained within the defined cylinder.
	
	To avoid biases from arbitrarily choosing a solar location in each galaxy, we created 360 all sky maps for each simulation, varying the azimuthal angle of the solar position. All positions were at a galactocentric radius of 8~kpc. At each of these locations, we built up a HEALPix array of unit vectors (with NSIDE~=~128) and integrated the electron density along each sightline out to the virial radius. The integration was performed using the \texttt{vortrace} code\footnote{\url{https://github.com/gusbeane/vortrace}} which uses the base Voronoi mesh to compute the ray-tracing (as opposed to a kernel-based approximation). This algorithm is similar to the one implemented by \cite{konietzka2025} for reconstructing the traversed line segments within the underlying Voronoi mesh.

	\section{All-Sky Dispersion Measures} \label{sec:dms}
	
	\begin{deluxetable}{llcccc}
		\tablecaption{Median DM, mean DM, standard deviations, and 16th--84th percentile scatter across the sky.\label{tab:dm}}
		\tablehead{\colhead{Galaxy} & \colhead{Resolution} & \colhead{Median} & \colhead{Mean} & \colhead{Std Dev} & \colhead{$16-84\%$}}
		\startdata
		Au6 &  &  &  &  &  \\
		& Default & $29.6\pm0.7$ & $40.9\pm0.8$ & $33.5$ & $17.1 - 64.4$ \\
		& 1 kpc & $24.9\pm1.2$ & $39.0\pm1.1$ & $37.8$ & $14.3 - 63.1$ \\
		& 500 pc & $18.9\pm0.2$ & $28.5\pm0.6$ & $29.7$ & $6.5 - 47.8$ \\
		& 200 pc & $26.8\pm0.8$ & $36.6\pm0.3$ & $32.9$ & $13.9 - 54.9$ \\
		Au8 &  &  &  &  &  \\
		& Default & $39.4\pm1.6$ & $54.0\pm1.5$ & $42.9$ & $22.3 - 83.3$ \\
		& 1 kpc & $27.7\pm1.2$ & $41.5\pm1.8$ & $38.7$ & $15.7 - 67.1$ \\
		& 500 pc & $25.7\pm0.5$ & $35.1\pm0.6$ & $32.7$ & $9.9 - 56.4$ \\
		TNG-A &  &  &  &  &  \\
		& Default & $35.1\pm0.7$ & $42.7\pm0.8$ & $30.4$ & $18.2 - 64.2$ \\
		& 1 kpc & $28.9\pm0.6$ & $38.9\pm0.5$ & $30.9$ & $17.2 - 59.5$ \\
		& 500 pc & $26.0\pm1.0$ & $34.2\pm1.3$ & $29.8$ & $9.5 - 58.6$ \\
		& 200 pc & $37.2\pm1.4$ & $42.4\pm0.9$ & $32.3$ & $11.4 - 69.3$ \\
		TNG-B &  &  &  &  &  \\
		& Default & $35.5\pm1.0$ & $42.8\pm0.7$ & $27.1$ & $22.6 - 60.1$ \\
		& 1 kpc & $26.9\pm0.3$ & $31.7\pm0.4$ & $19.9$ & $18.4 - 41.2$ \\
		& 500 pc & $33.7\pm0.2$ & $37.8\pm0.3$ & $17.6$ & $24.4 - 51.1$ \\\hline
		YT20 &  & $37.8$ & $47.6$ & $27.0$ & $31.3 - 61.2$ \\
		\enddata
		\tablecomments{All values are taken as the median across all 360 solar positions for each simulation. The uncertainty on the mean is the standard deviation of the 360 all-sky mean DM values at different solar positions. All values are in\dmunits.}
	\end{deluxetable}
	
	\begin{figure}[t]
		\centering
		\includegraphics[width=0.44\textwidth]{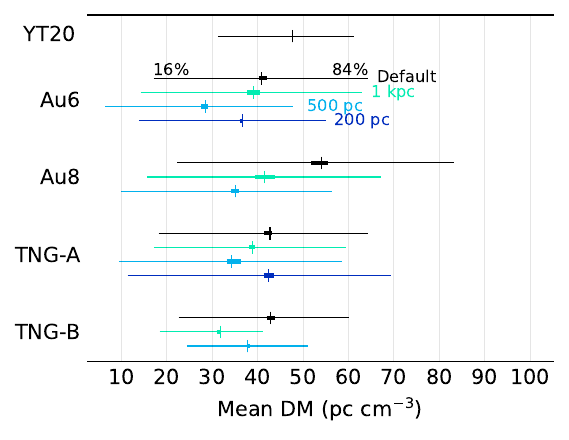}
		\caption{The distribution of DM values for each of the simulations compared with \yt. Each horizontal bar represents a simulation where they are grouped by IC and colored by resolution. The full extent of the bars represent the median $16-84$\% range across the 360 solar locations. The inner thicker bar with the vertical line represents the 16, 50, and 84\% values of the mean DM across the whole sky from the sample of 360 solar locations. These values are compared against the mean (vertical line) and $16-84$\% range (extent of line) of the \yt\ model shown above the axis. There is relatively little variation based on solar position, but there is significant variation from sightline to sightline.}
		\label{fig:meandms}
	\end{figure}
	
	With the \texttt{vortrace} ray-tracing method, we obtained all-sky maps of integrated dispersion measure for 360 different solar locations for each of the 14 simulations. An example of this data is shown in Figure~\ref{fig:allsky} with the all-sky map on the top, colored by DM, where the left half of the image shows the Default resolution simulation for Au6, while the right half shows the highest resolution 200~pc run for Au6. The bottom histograms show the distribution of DM values for all the pixels in the all-sky images for the two resolutions. The minimum, mean, and maximum values are shown as vertical lines (and labeled on the $x$-axis).
	At higher resolution, we see the minimum and mean values drop and the maximum value increase $-$ indicating more low DM sightlines as well as greater anisotropy.
	% While the mean values are the same between the two resolutions, we can see that the full extent of DM values is extended from $8-396$ to $5-445$.
	This is also visible qualitatively in the all-sky map in which there are more low-DM regions in the high resolution simulation due to the increased substructure that we are able to resolve.
	
	\begin{figure*}[t]
		\centering
		\includegraphics[width=1.0\textwidth]{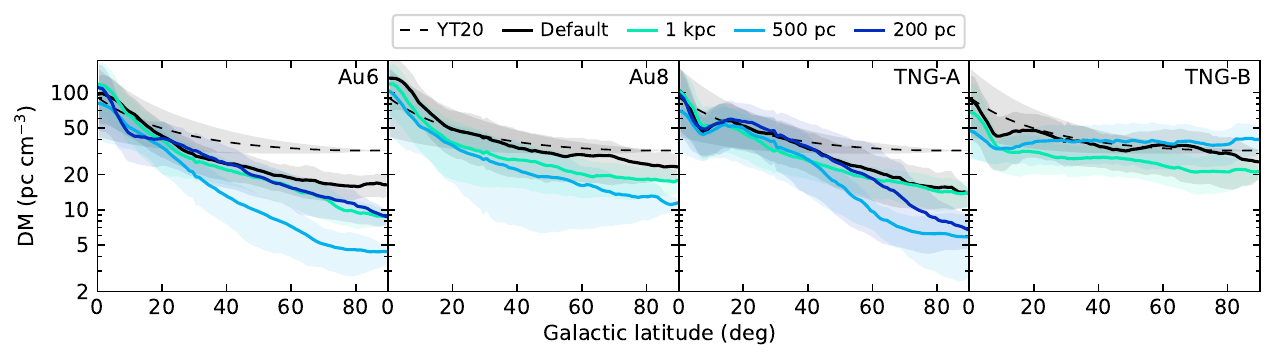}
		\caption{DM as a function of galactic latitude for the simulations, compared against \yt. Each panel shows a different galaxy, and each colored line is a different resolution. The solid lines are generated by taking the median over the 360 solar positions of the mean DM within each latitude bin. The shaded regions show the solar position median of the 16th and 84th percentile within each latitude bin. The dashed line represents the mean value from \yt, with the grey shaded region denoting the 16--84\% range of values for a given latitude bin.}
		\label{fig:dmvsb}
	\end{figure*}
	
	For these subsequent figures and results, we are reporting summaries based on calculations across all 360 all-sky images, each from a different solar location in the simulation. In general, we have attempted to calculate properties for each of the all-sky maps individually, and then report the median values over the 360 different solar locations. We feel that this provides the best comparison to the models and observations in which we a single all-sky DM map for our solar location.
	
	In Figure~\ref{fig:meandms}, we show the mean all-sky DM and sightline to sightline variation for all 14 simulations. Each horizontal line denotes a different simulation $-$ grouped by galaxy and colored by resolution. The full extent of the lines denote the 16--84\% range of DM values (we computed the 16th and 84th quartile for each of the 360 all-sky HEALPix images and are reporting the median of those 360 values). The vertical tick shows the mean all-sky DM value (the median of all the 360 all-sky mean DM values). Finally, the thick horizontal bar represents the 16--84\% range in the mean all-sky DM value across the 360 solar locations. This shows us that the mean DM values are relatively stable throughout the galaxy (the thick horizontal bars are small), while there is significant spread in DM values from sightline to sightline (the full extent of the thin lines is large). These simulation results are then compared against the 16--84\% range and mean all-sky DM value calculated with the \yt\ model. Note that since the \yt\ model is spherically symmetric, there is no variation in the mean all-sky DM value as a function of solar location. In general, the \yt\ mean DM value is higher than the mean values from the simulations, and the 16--84\% range from \yt\ is smaller, especially on the low end. Looking at the full range of values, \yt\ extends from $29-249$\dmunits, while the simulations extend down well below 10\dmunits, and up above 400\dmunits (see, e.g. Figure~\ref{fig:allsky}).
	
	Table~\ref{tab:dm} lists the all-sky median and mean DM values as well as the standard deviation and 16th and 84th percentiles. As above, these values are calculated individually for each all-sky image and then the values reported are the medians across the 360 solar locations. The uncertainties quoted for the median and mean values arise from the standard deviation of these values across the 360 solar locations. In all simulations except the default resolution of Au8, both the median and the mean all-sky DM values are lower than the values from \yt. Furthermore, the standard deviation of the all-sky images is larger than \yt\ for all simulations except TNG-B.
	
	While this is comparing the all-sky properties, Figure~\ref{fig:dmvsb} shows the DM distribution as a function of Galactic latitude. Each panel shows a different galaxy, and again, each color represents a different resolution. The lines represent the mean DM as a function of $b$ (again taking the median across the 360 solar locations). The shaded regions denote the 16--84\% ranges in each $b$-bin. For \yt, we also show the mean DM and 16--84\% range as a function of $b$. While the DM values from the simulations and the \yt\ model are comparable for low $b$, in all but TNG-B, the high latitude DM values are significantly lower in the simulations, with significantly more variation.
	
	\subsection{Functional Model} \label{sec:model}
	
	For applications to interpreting extragalactic FRB detections, we provide in this section an all-sky map of MW halo DM as a function of $\ell$ and $b$. For each of our two highest resolution simulations, Au6 and TNG-A at 200~pc, we have all-sky DM maps for 360 different solar locations throughout the disk. By taking the median value for each pixel across all 360 maps, we obtain an average all-sky image, marginalizing over the specific solar position within the simulation. We can also obtain 16--84\% DM ranges for each pixel to estimate an uncertainty in our predicted DM for each sightline. Finally, by averaging the maps from the two galaxies, we can marginalize over specific evolutionary histories and isolated structures.
	
\begin{figure*}
		\centering
		\includegraphics[width=1.0\textwidth]{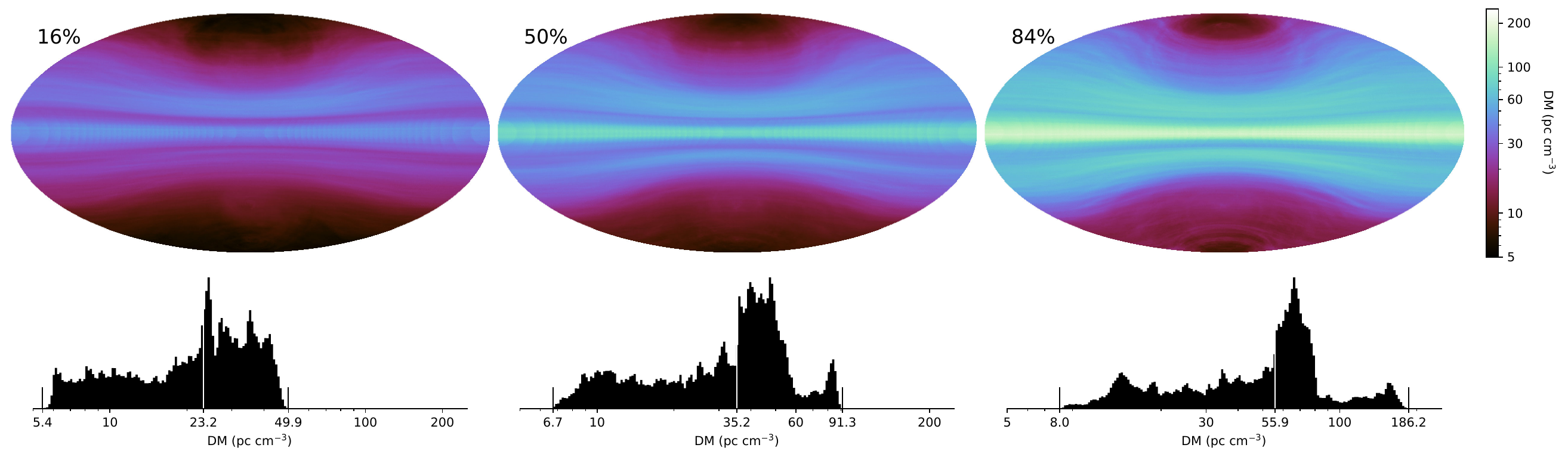}
		\caption{All-sky maps and histograms of MW halo DM for the combined high resolution model (azimuthally averaged combination of 200~pc Au6 and TNG-A simulations). Each pixel contains a distribution of DM values collated from the different solar locations as well as the different simulations. The different columns show different quantiles of those distributions within each pixel: 16, 50, and 84\% from left to right. The histograms show the distributions of pixel values with the minimum and maximum marked with vertical black lines (and labeled), and the mean shown as a vertical white line (and labeled).}
		\label{fig:quants}
	\end{figure*}

	Figure~\ref{fig:quants} shows the all-sky maps for the 16th, 50th (median), and 84th percentiles values for each pixel, averaged between the Au6 and TNG-A simulations at $z=0$. The horizontal stripes are a result of each all-sky image being rotated such that the galactic center lies at $\ell=0$. Thus coherent structures that are visible from multiple solar locations become smeared out across the sky. With the 16th and 84th percentile maps, we are able to assess the uncertainty in a given DM value at that sky position. % This is discussed further in Sections~\ref{sec:model} and \ref{sec:disc}.
	
	The uncertainty in these maps is driven slightly more by the sightline-to-sightline variation while the galaxy-to-galaxy variation is secondary. For Au6 and TNG-A, we measured the sightline-to-sightline variation by taking the standard deviation of all the DM values for each pixel across the 360 solar positions. To compare the two galaxies, we took the difference in median pixel values (taking the median over the 360 solar positions) between Au6 and TNG-A. Taking the mean across the whole sky of these different metrics shows that the average standard deviation for each galaxy individually is $\sim18$\dmunits\ while the average difference in median pixel value between the galaxies is 13\dmunits. Note that this standard deviation is measuring the distribution in values for a given pixel, while the std numbers in Table~\ref{tab:dm} represent the distribution across the sky for a given solar location.
	
	\begin{figure*}
		\centering
		\includegraphics[width=1.0\textwidth]{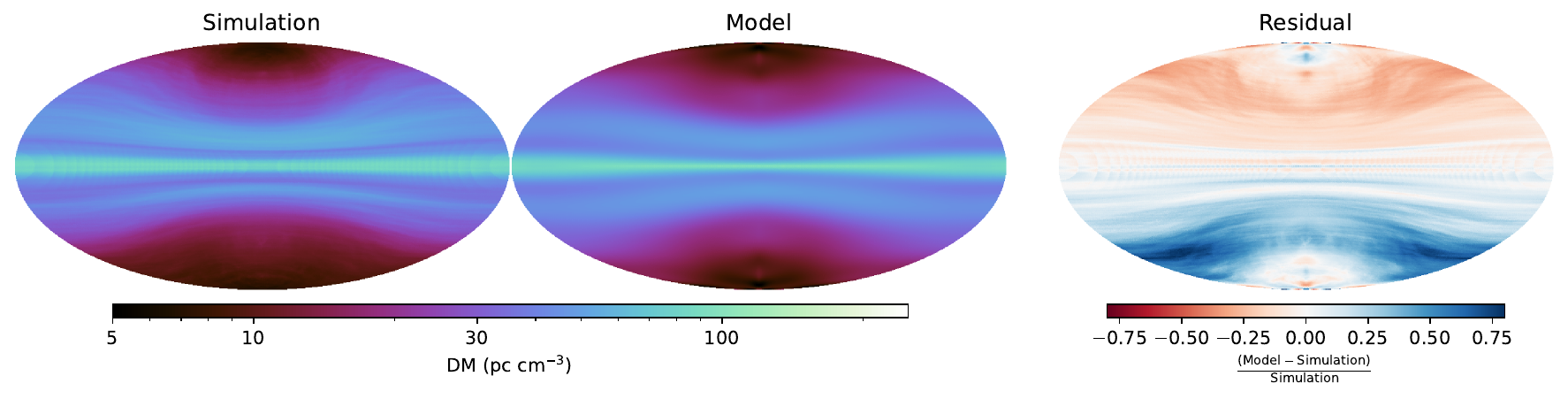}
		\caption{All-sky maps of our combined high resolution model (left; this panel is the same as the center top panel in Figure~\ref{fig:quants}) compared with the polynomial fit model (center). The right panel shows the residual (percent error; (model $-$ simulation)/simulation). The residuals are mainly due to a north-south asymmetry in the simulation that is not included in the model (we are fitting to $|b|$).}
		\label{fig:fit}
	\end{figure*}
	
	Finally, we fit these maps with polynomials with similar structure as in \yt:
	\begin{equation}
		\mathrm{DM}(\ell,b) = \sum_{i+j\leq0}^n c_{ij}|\ell|^i|b|^j.
	\end{equation}
	The $c_{ij}$ coefficients are shown in Table~\ref{tab:coefs} (if $\ell$ and $b$ are given in radians), published in full as a machine-readable table online. Polynomials of order 3 to 20 were used and the selected best-fit model (of order 10) was the lowest order that had a root mean square error (RMSE) within one standard error of the best model. Figure~\ref{fig:fit} shows the simulation DM on the left (this is the same as the center panel of Figure~\ref{fig:quants}), the polynomial fit in the center panel, and the fractional difference between the simulation and the fit on the right: (model - simulation)/simulation. Note that the polynomial is fit to $|\ell|$ and $|b|$ and therefore is symmetric vertically and horizontally. The main discrepancies between the simulation and the model are due to a north$-$south asymmetry which is why a higher order fit does not improve the RMSE.
	
\begin{figure}
		\centering
		\includegraphics[width=0.44\textwidth]{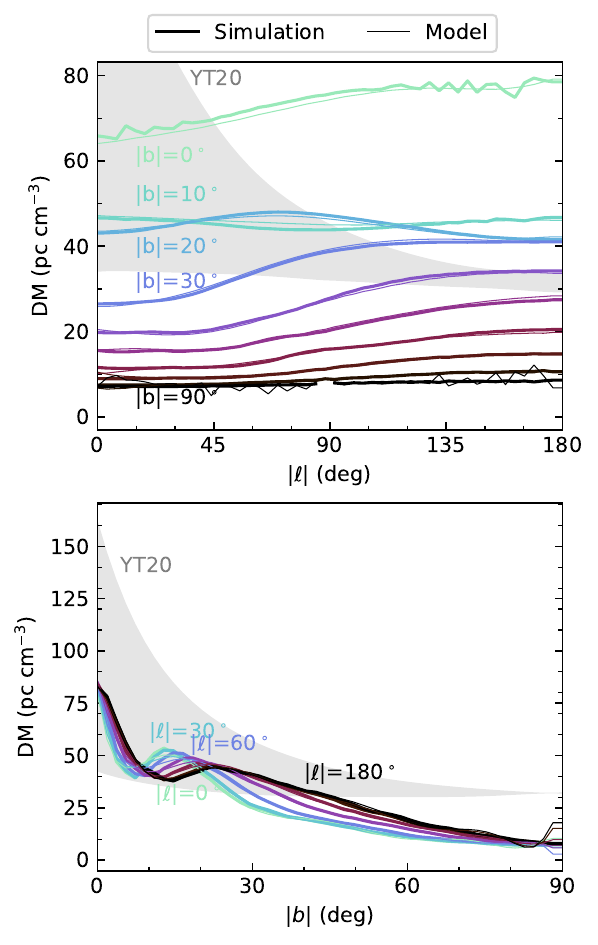}
		\caption{The MW halo DM value from the simulation (thick lines) compared with the polynomial model (thin lines) and the \yt\ model (grey shaded regions). The top panel shows the DM as a function of Galactic longitude for different bins in latitude (different colors). Each line is separated by $10^\circ$ in $b$ and comprises the average value within a bin of width $10^\circ$. The bottom panel shows the comparable relationships in the other dimension: DM vs latitude for varying longitude values. In this case, we use longitude values separated by $30^\circ$.}
		\label{fig:fit_lb}
	\end{figure}

	Figure~\ref{fig:fit_lb} also shows the comparison between the simulations (thick lines) and polynomial fit (thin lines) as a function of $\ell$ and $b$. In the top panel, each curve shows the DM vs. $\ell$ for a different $b$ (averaged within $\Delta b=10^\circ$ bins), and the bottom panel shows DM vs. $b$ for different $\ell$ (averaged within bins of $\Delta \ell=30^\circ$). Interestingly, we see that the DM increases towards the galactic anticenter ($\ell=180^\circ$). This is because we have excised the disk of the simulation (a cylindrical region with radius $\sim25$~kpc and height 10~kpc) and the solar location is offset from the center of the galaxy. The sightlines toward the galactic anticenter pass through more material due to the fact that the vertical edge of the cylinder is closer to the solar position in that direction. The bump in the DM vs. $b$ curves (around $b=20^\circ$) is also a remnant of the cylinder cut out and constitute the relatively high density disk-halo interface material just outside the excised cylinder.
	
	For ease of applicability to future observations, we have made these results accessible via a \texttt{pip} installable python package accessible via github\footnote{\gitlink}. This module provides the HEALPix maps used to make Figures~\ref{fig:quants} and \ref{fig:fit} in addition to the polynomial fit coefficients and information for estimation of the MW halo DM as a function of $\ell$ and $b$. In addition to the fit for the median map (shown in Figures~\ref{fig:fit} and \ref{fig:fit_lb}), we also provide order 10 fits to the 16\% and 84\% maps. Thus, with the \texttt{fit\_dm($\ell$,$b$,with\_err=True)} function, one can obtain the MW halo DM estimate at location ($\ell$, $b$) as well as the 16--84\% range as an uncertainty on the value. This has not been possible with previous models and is a crucial element for statistical analysis of FRB DMs.
	
	\section{Comparison to Observables} \label{sec:obsv}
	
	\begin{deluxetable}{ccccc}
		\tablecaption{Polynomial coefficients $c_{ij}$}
		\tablehead{$i$ & $j=0$ & $j=1$ & $j=2$ & $\cdots$}
		\startdata
		$0$ & $100.797$ & $-1417.15$ & $12566.5$ & \\
		$1$ & $-6.46523$ & $522.18$ & $-7772.41$ & \\
		$2$ & $17.3267$ & $-13.8052$ & $128.216$ & \\
		\vdots &  &  & &
		\enddata
		\tablecomments{Model: $f(l,b)=\sum_{i+j\leq 10} c_{ij} |\ell|^i |b|^j$. Table~\ref{tab:coefs} is published in its entirety in machine-readable format with full precision. Alternatively, the fit coefficients are available via the github package: \gitlink.}
		\digitalasset
		\label{tab:coefs}
	\end{deluxetable}
	
	% In addition to providing the direct results of the simulations above, we acknowledge that these simulations are not perfect mat
	While these systems have been specifically selected as MW analogs, and we expect that the halo contribution to the DM to be statistically similar to our own Galaxy, these simulations are not perfect reproductions of the MW. Therefore, in addition to the median and uncertainty maps provided above, we explore correlations between MW halo DM and other observable properties. Since sightline to sightline variation is one of the biggest drivers of uncertainty in the DM value, being able to estimate the halo DM contribution based on other observables along the same line of sight could prove fruitful.
	% These simulations have been specifically selected as Milky Way analog systems. However, it is clear that sightline to sightline variation is one of the biggest drivers of uncertainty in DM value. In order to predict the Milky Way halo DM contribution across the full sky, we need to calibrate models using these simulations that we can then apply to our own Galaxy. Simply taking the DM values from the simulation is a good start, but we can do better.
	
\begin{figure}[t]
		\centering
		\includegraphics[width=0.44\textwidth]{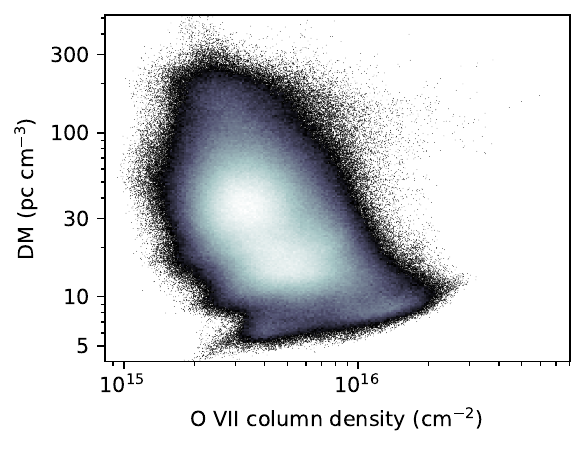}
		\caption{DM as a function of \ovii\ column density for the 200~pc resolution run of Au6. The histogram (and underlying points) are a random selection of 2 million sightlines from all of the 360 solar position all-sky maps. The \ovii\ column densities are computed in the same manner as the DMs (see Section~\ref{sec:methods}) including the cylinder disk cutout.}
		\label{fig:ovii}
	\end{figure}

	Interestingly, \ovii\ column density does not seem to trace the halo DM at the sightline level. Figure~\ref{fig:ovii} shows the relation between DM and \ovii\ column density sightline by sightline. The \ovii\ column densities were calculated using \textsc{trident}\footnote{\url{https://github.com/trident-project/trident}} to compute the ion fractions \citep{hummels17} and \texttt{vortrace} to integrate along the sightlines. To populate this histogram, we have taken a random sampling of 2 million pixels across the 360 all-sky maps and plotted the matched DM value and the \ovii\ column density for each sightline. Note that both of these values were calculated with the cylinder disk cutout. Clearly there is no trend here. This is mildly concerning as the \yt\ model was calibrated based on \ovii\ data in the MW. We do note that we are simply computing the \ovii\ column density, not the emissivity, so this relation could be different in the case of comparing against the local \ovii\ emission.
	
	Even more than a lack of correlation, Figure~\ref{fig:ovii} shows an anticorrelation between DM and \ovii\ column density. We believe this is due to the fact that the \ovii\ column density is highest at the poles since this is where the hottest gas lives. However, in these directions, the total gas density is quite low and thus the DM is correspondingly low. In other words, the DM is driven by total gas density, and the \ovii\ column density is more driven by metallicity and temperature (such that the \ovii\ state is populated).
	
	% Luckily, the simulations do reveal a couple observables that do track DM on a sightline to sightline level relatively well. First, we have neutral hydrogen, \hi. The top panel of Figure~\ref{fig:hi}, analagous to Figure~\ref{fig:ovii}, shows the integrated halo DM as a function of \hi\ column density with the cylinder disk cutout (``CGM Only''). Here, there is a clear trend, however the scatter is quite large especially at higher column densities. Most of the sightlines lie in the $10^{16}-10^{18}$~cm$^{-2}$ column density regime which is not currently accessible in all-sky high-velocity \hi\ maps \citep{westmeier18}. The bottom panel shows the total \hi\ column integrated from $r=0$ out to the virial radius (``Full Sightline''). Due to the local interstellar medium contribution, the \hi\ column values are greatly compressed to $n_\mathrm{HI}\gtrsim10^{20}$~cm$^{-2}$, and thus the scatter is even further increased. So while \hi\ provides a local DM scaling relation in theory, due to observational constraints, it is not particularly predictive.
	
	% \begin{figure}[t]
		%     \centering
		%     \includegraphics[width=0.44\textwidth]{DM_vs_HI.pdf}
		%     \caption{DM as a function of \hi\ column density across all sightlines for a given solar position of the 200~pc resolution run of Au6. The top panel uses the cylinder disk cutout to integrate the \hi\ column density, while the bottom panel integrates from $r=0$. Note that the $x$-axes have different scales.}
		%     \label{fig:hi}
		% \end{figure}
	
\begin{figure}[t]
		\centering
		\includegraphics[width=0.44\textwidth]{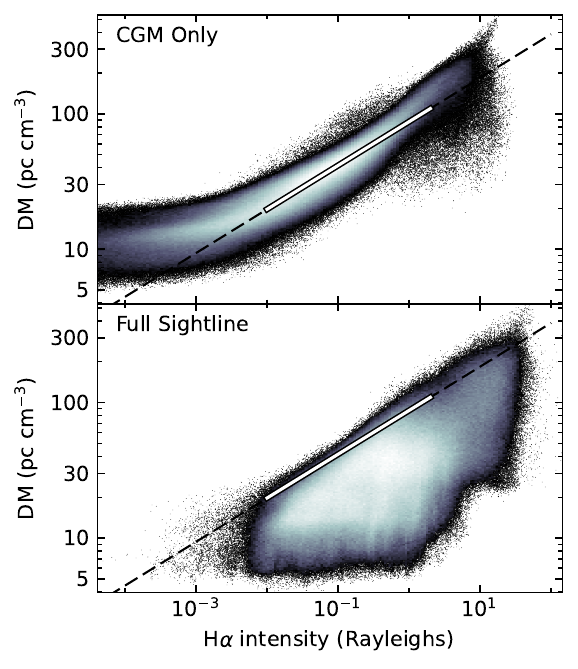}
		\caption{DM as a function of \ha\ intensity for the 200~pc resolution run of Au6. As in Figure~\ref{fig:ovii}, the data are a random selection of 2 million sightlines from all of the 360 solar position all-sky maps. The top panel uses the cylinder disk cutout to integrate the \ha\ intensity, while the bottom panel integrates from $r=0$.}
		\label{fig:ha}
	\end{figure}

	Luckily, the simulations do reveal an observable that tracks DM on a sightline to sightline level relatively well: H$\alpha$. The top panel of Figure~\ref{fig:ha}, analagous to Figure~\ref{fig:ovii}, shows the integrated halo DM as a function of \ha\ intensity with the cylinder disk cutout (``CGM Only''). Here there is a clear trend and we have fit a power law relation within the region $10^{-2}<I_\mathrm{H\alpha}<2$~Rayleighs which is shown as the white line and corresponds to $\log_{10}(\mathrm{DM})=0.32\times \log_{10}(I_\mathrm{H\alpha}/R)+1.93$, where $I_\mathrm{H\alpha}/R$ is the intensity in Rayleighs. The relation is extended beyond this range with a dashed line. This same line appears on the bottom panel which shows the total \ha\ intensity integrated from $r=0$ out to the virial radius (``Full Sightline''). Due to the local interstellar medium contribution, the \ha\ intensities and the scatter are increased. We do see, however, that the linear fit from the CGM Only data provides an upper limit for DM as a function of \ha\ intensity. Additionally, since we do not resolve \hii\ regions in the simulation, we would expect the actual \ha\ intensity to be higher. Thus, this fit is even more robustly an upper limit to the DM as a function of $I_\mathrm{H\alpha}$.
	
\begin{figure*}
		\centering
		\includegraphics[width=1.0\textwidth]{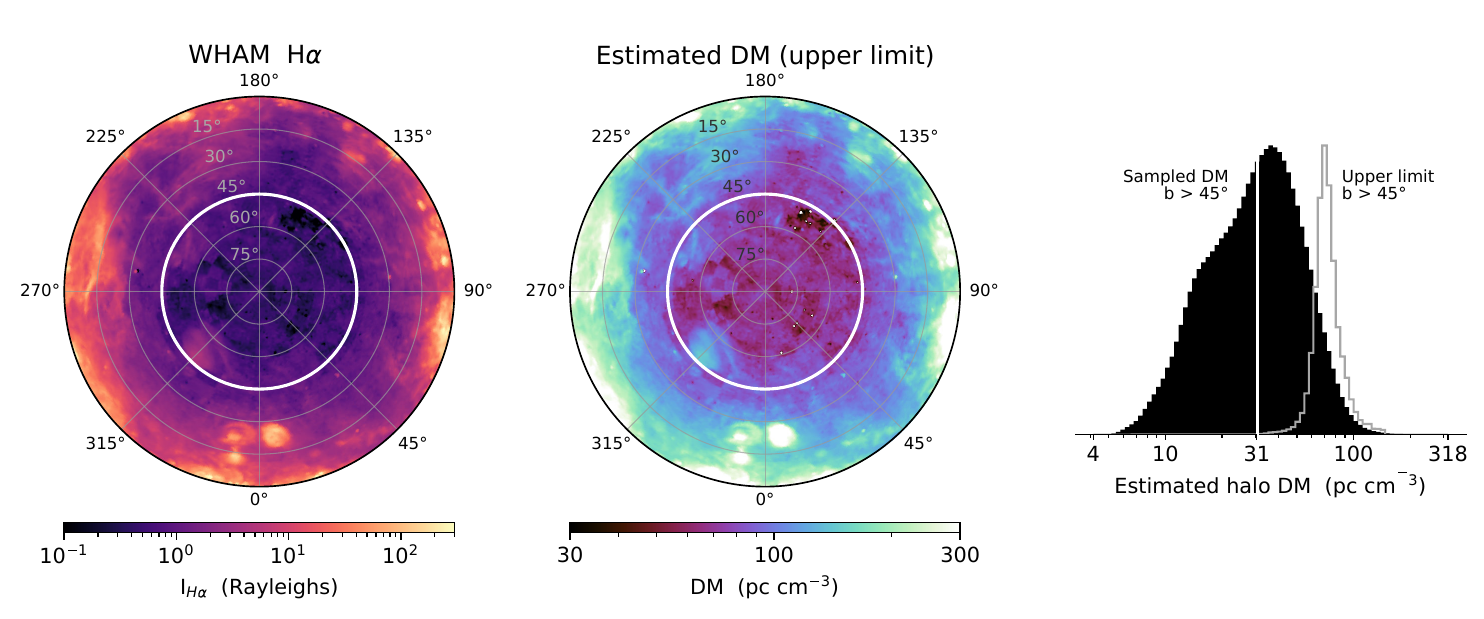}
		\caption{DM values for the Northern Hemisphere extrapolated from WHAM \ha\ observations \citep{wham}. The WHAM data are shown on the left, and the center panel shows the upper limit of the MW halo DM calculated from the linear fit in Figure~\ref{fig:ha}. The right panel shows DM distributions for the $b>45^\circ$ pixels in two cases: the grey, unfilled distribution corresponds to the upper limit values in the center panel; and the black, solid distribution comes from a Monte Carlo sampling of DM values from the distributions in Figure~\ref{fig:ha}.}
		\label{fig:wham}
	\end{figure*}

	Figure~\ref{fig:wham} now shows the implied MW halo DM calculated from the Northern Hemisphere of the WHAM H$\alpha$ data \citep{wham}. The left panel shows the WHAM data and the center panel shows the estimated DM for each pixel calculated using the linear fit shown in Figure~\ref{fig:ha}. Since the WHAM data include all emission down to $r=0$, Figure~\ref{fig:ha} shows us that this linear fit is an upper limit for the DM given an \ha\ intensity (Figure~\ref{fig:ha} bottom panel). In the right panel of Figure~\ref{fig:wham}, we have produced two histograms for the DM values for the $b>45^\circ$ data: the grey, unfilled distribution is the sky area-weighted histogram of the upper limit values from the center panel image; the black, solid histogram shows the sky area-weighted distribution of Monte Carlo sampled DM values for each given WHAM \ha\ pixel. Taking the bottom panel of Figure~\ref{fig:ha}, we sample 200 DM values for each \ha\ intensity value in the WHAM data from the distribution of DMs at that \ha\ intensity in the simulation. This necessarily produces lower DMs since the linear fit is an upper limit, and we see that while the median is still at 30\dmunits, there is a non-negligible portion of the distribution with DM values less than 10\dmunits. As stated above, due to the unresolved nature of the ISM in the simulation, even this Monte Carlo sampling can be considered an upper limit on DM along each sightline. This is consistent with our findings from the simulation that certain sightlines can have very low MW halo DM contributions.
	
	\section{Discussion \& Conclusion} \label{sec:disc}
	
	In this work, we have investigated the MW halo contribution to FRB DMs in high resolution cosmological zoom-in simulations.
	We find that the all-sky mean DM values from the simulations are, in general, slightly lower than previous models, with the lower limit of possible DM contributions being much lower.
	% While the all-sky mean DM values are broadly consistent with previous models (\yt), the lower limit of possible DM contributions is much lower.
	Sightlines towards the galactic poles can have MW halo DM values $<10$\dmunits. We have also shown that within the IllustrisTNG galaxy formation model, \ovii\ column densities are not good tracers of DM at the sightline to sightline level. This brings into question the applicability of the \yt\ model since it was derived assuming an \ovii\ to DM relation. \ha, on the other hand, does provide a good sightline to sightline estimate of the halo DM, however local ISM features make this extrapolation difficult. We have used the WHAM \ha\ map to corroborate our prediction that the MW halo DM contribution can be very low depending on the sightline by mapping \ha\ intensity to halo DM statistically from our simulation to the real MW. We do see sightlines with predicted DM values $<10$\dmunits.
	
\begin{figure}[t]
		\centering
		\includegraphics[width=0.44\textwidth]{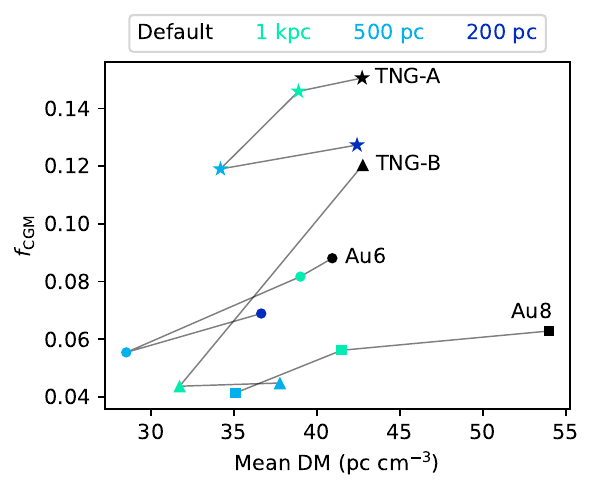}
		\caption{CGM gas fraction vs mean DM for the different galaxies at different resolutions. The different marker styles denote the different galaxies, while the different marker colors denote the different CGM resolutions. Results for galaxies at different resolutions are connected by black lines.}
		\label{fig:fcgm}
	\end{figure}

	In addition to the correlations with gas tracers at the sightline level, we have explored how the mean all-sky DM tracks CGM gas fraction in the simulations.
	% At the level of the mean all-sky DM, our simulations show that CGM gas fraction does not actually correlate with mean DM. 
	Figure~\ref{fig:fcgm} shows $f_\mathrm{CGM}=M_\mathrm{CGM}/(0.16\times M_{200})$ vs mean all-sky DM (as calculated in Figure~\ref{fig:meandms} and listed in Table~\ref{tab:dm}), where $M_\mathrm{CGM}$ is the total gas mass within the virial radius of the galaxy excluding the cylinder disk cutout used in the DM integration, and $M_{200}$ is the virial mass of the galaxy. Each marker style shows the values for a different galaxy, while the colors represent the different resolution simulation runs. Runs with the same galaxy at different resolutions are connected by black lines. Here, we can see that, with the exception of Au8-Default, all the mean all-sky DM values are $\sim30-40$\dmunits, while $f_\mathrm{CGM}$ ranges from 4\% up to 15\%, with no clear trend between the two. Thus, CGM gas fraction is not a good tracer for mean DM of the halo, at least when viewing from the inside. This relation may improve when considering external galaxies.
	
	This lack of correlation could be due to the phase distribution of the gas. DM is tracing free electrons, while there is a significant portion of the CGM mass in neutral material. In our simulations, \hi\ accounts for 30--50\% of the total gas mass in the CGM, thus weakening the power of DM to measure $f_\mathrm{CGM}$.
	
	The main data product that we are providing with this work is a Python package that provides simple API access to the simulated all-sky DM map as well as a polynomial model fit. This package can be used to estimate the MW halo DM contribution for any Galactic $\ell$ and $b$ directly from the simulations (averaged across different solar positions and the two highest resolution simulations), or from the polynomial model fit. Moreover, the 16--84\% ranges of DM values for different solar positions are provided within the API. This provides, for the first time, a measurement of the uncertainty in MW halo DM for any given sightline. This will be a critical piece of future FRB analysis in order to better understand the potential implications of various observations.
	
\begin{figure}[t]
		\centering
		\includegraphics[width=0.44\textwidth]{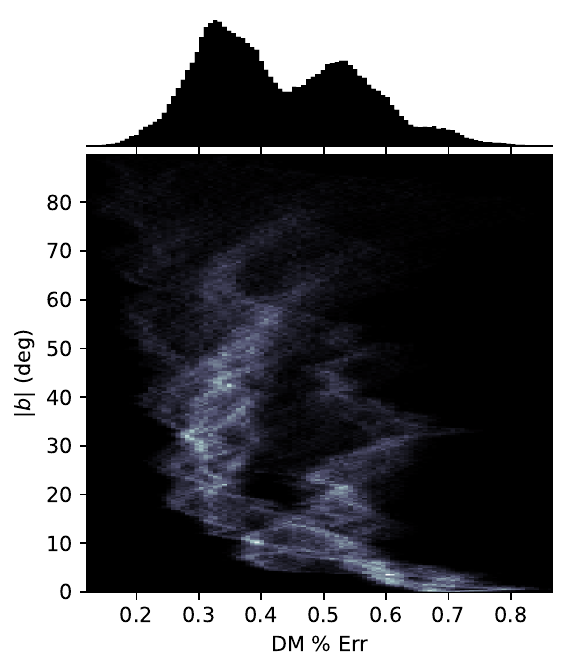}
		\caption{Distribution of DM uncertainties as a function of Galactic latitude, $b$. The histogram on top shows the distribution of \% error on DM ($\rm (DM_{84}-DM_{16})/2/DM_{50}$) with a peaks at 0.33 and 0.53. The peak at lower \% error corresponds to higher latitudes, while the higher \% error corresponds to lower latitudes.}
		\label{fig:errhist}
	\end{figure}

	This quantification of the uncertainty on the MW halo DM is important. In this work, we are able to obtain an estimate for this by averaging the all-sky simulated DM values across 360 solar positions within the galaxy (all around the solar circle at $r=8$~kpc). The distribution of fractional uncertainties across all the pixels ($(\text{DM}_{84} - \text{DM}_{16})/2/\text{DM}_{50}$) is bimodal with a peak at 0.33 and 0.53. Figure~\ref{fig:errhist} shows this distribution (on the $x$-axis) compared against the Galactic latitude of the pixels (on the $y$-axis). The histogram on top is simply the distribution of fractional uncertainties.
	The peak at 0.33 contains 57\% of the pixels, and mostly consists of intermediate and high latitudes. The peak at 0.53 is dominated by the mid- and low-latitude regions, while the disk ($|b|\lesssim5^\circ$) has even higher fractional errors at $\sim0.7$. These uncertainties can be as low as a few \dmunits towards the poles (due to their lower $\rm DM_{50}$ values), however the uncertainties towards the disk peak at 20\dmunits, and extend up to $\sim60$\dmunits. Thus, they must be taken into consideration when analyzing future FRB observations.
	
	An additional benefit of the ENGAWA simulation suite is the multiple different resolution levels. For each galaxy we have fixed spatial resolution in the CGM down to 200~pc. While there are not dramatic changes with resolution, we generally see more uniform distributions of DM values as resolution increases. This is shown in the histograms at the bottom of Figure~\ref{fig:allsky}. The mean all-sky DM values also generally decrease, with the lower bound of possible DMs decreasing as well. The 500~pc simulations are often lower than the 200~pc simulations and we believe this to be a difference in the gas distribution for this particular simulation resolution. While we do not see converged behavior yet with these resolutions, there is still significant stochasticity associated with each halo. A larger sample of galaxies would help disentangle resolution convergence from galaxy evolution.
	
	Recently, \citep{huang25} investigated the DM contribution from the local universe in the \textsc{Hestia} simulations. While their analysis extended out to 120~Mpc, they do include the DM contribution from the circumgalactic gas of their simulated MW-like galaxies. They used a similar technique to excise the galactic disks (cylinder cutouts) and they find similar results with an average all-sky DM of 45\dmunits\ with a standard deviation of 17\dmunits\ (excluding the Magellanic Clouds). While their average values are consistent with our findings, the higher resolution simulations presented here have sightlines with even lower DMs, down to 3--8\dmunits, as mentioned above. Thus the standard deviations of the all-sky DMs in our simulations are larger ($\sim30-40$\dmunits). This highlights the importance of resolution in probing the lower limits of MW halo DM. We see a similar result in comparison with the \yt\ model. The mean all-sky DM values are slightly lower, but the lower end of the DM distributions is much more extended in our simulations due to the self-consistent anisotropy from the cosmological simulation.
	
	An additional consideration for the halo of our Milky Way is the Magellanic Clouds. This is not taken into account in our models, so we would expect the total DM from halo gas in the Southern hemisphere to be higher due to the contribution from the LMC, SMC, and Magellanic Stream \citep{donghia16,lucchini20,dk22}. This is considered in \citet{huang25}, and they estimate a contribution of up to $\sim40$\dmunits.
	
	Our conclusions are as follows:
	\begin{enumerate}
		\item We derive a new model for the MW halo contribution to FRB DM from high resolution cosmological zoom-in simulations.
		\item The all-sky mean DM contribution from the MW halo from high-resolution simulations is generally slightly lower than previous models.
		\item Simulations predict much more sightline to sightline variation than previous models with DM values $<10$\dmunits\ towards the poles.
		\item \ha\ intensity correlates well with MW halo DM, while \ovii\ column density does not.
		\item $f_\mathrm{CGM}$ does not correlate with all-sky mean DM in this model.
		\item We have provided a Python package with API access to the MW halo DM values from the simulation (as well as a polynomial model fit) as a function of Galactic longitude and latitude. This package also includes 16 and 84\% maps to estimate the uncertainty towards any given sky direction.
	\end{enumerate}
	
	\begin{acknowledgements}
		The computations in this paper were run on the FASRC cluster supported by the FAS Division of Science Research Computing Group at Harvard University.
		Support for SL was provided by Harvard University through the Institute for Theory and Computation Fellowship. L.C. and S.M. acknowledge support from the U.S. National Science Foundation Astronomy and Astrophysics Research Grants program under award AST-2508734.
		Some of the results in this paper have been derived using the healpy and HEALPix package.
	\end{acknowledgements}
	
	\software{matplotlib v3.10.0 \citep{hunter07}, numpy \citep{harris20}, healpy \citep{healpy}, \textsc{trident} v1.3 \citep{hummels17}, \textit{yt} v4.5 \citep{turk11}, vortrace (\url{https://github.com/gusbeane/vortrace})}
	
	\bibliography{references}{}

@ARTICLE{lorimer2007,
       author = {{Lorimer}, D.~R. and {Bailes}, M. and {McLaughlin}, M.~A. and {Narkevic}, D.~J. and {Crawford}, F.},
        title = "{A Bright Millisecond Radio Burst of Extragalactic Origin}",
      journal = {Science},
         year = 2007,
        month = nov,
       volume = {318},
       number = {5851},
        pages = {777},
          doi = {10.1126/science.1147532},
archivePrefix = {arXiv},
       eprint = {0709.4301},
 primaryClass = {astro-ph},
       adsurl = {https://ui.adsabs.harvard.edu/abs/2007Sci...318..777L}
}

@ARTICLE{gupta12,
       author = {{Gupta}, A. and {Mathur}, S. and {Krongold}, Y. and {Nicastro}, F. and {Galeazzi}, M.},
        title = "{A Huge Reservoir of Ionized Gas around the Milky Way: Accounting for the Missing Mass?}",
      journal = {\apjl},
         year = 2012,
        month = sep,
       volume = {756},
       number = {1},
          eid = {L8},
        pages = {L8},
          doi = {10.1088/2041-8205/756/1/L8},
archivePrefix = {arXiv},
       eprint = {1205.5037},
 primaryClass = {astro-ph.HE},
       adsurl = {https://ui.adsabs.harvard.edu/abs/2012ApJ...756L...8G}
}

@ARTICLE{macquart2020,
       author = {{Macquart}, J.-P. and {Prochaska}, J.~X. and {McQuinn}, M. and {Bannister}, K.~W. and {Bhandari}, S. and {Day}, C.~K. and {Deller}, A.~T. and {Ekers}, R.~D. and {James}, C.~W. and {Marnoch}, L. and {Os{\l}owski}, S. and {Phillips}, C. and {Ryder}, S.~D. and {Scott}, D.~R. and {Shannon}, R.~M. and {Tejos}, N.},
        title = "{A census of baryons in the Universe from localized fast radio bursts}",
      journal = {\nat},
         year = 2020,
        month = may,
       volume = {581},
       number = {7809},
        pages = {391-395},
          doi = {10.1038/s41586-020-2300-2},
archivePrefix = {arXiv},
       eprint = {2005.13161},
 primaryClass = {astro-ph.CO},
       adsurl = {https://ui.adsabs.harvard.edu/abs/2020Natur.581..391M}
}

@ARTICLE{bland-hawthorn16,
       author = {{Bland-Hawthorn}, Joss and {Gerhard}, Ortwin},
        title = "{The Galaxy in Context: Structural, Kinematic, and Integrated Properties}",
      journal = {\araa},
         year = 2016,
        month = sep,
       volume = {54},
        pages = {529-596},
          doi = {10.1146/annurev-astro-081915-023441},
archivePrefix = {arXiv},
       eprint = {1602.07702},
 primaryClass = {astro-ph.GA},
       adsurl = {https://ui.adsabs.harvard.edu/abs/2016ARA&A..54..529B}
}

@ARTICLE{grcevich09,
       author = {{Grcevich}, Jana and {Putman}, Mary E.},
        title = "{H I in Local Group Dwarf Galaxies and Stripping by the Galactic Halo}",
      journal = {\apj},
         year = 2009,
        month = may,
       volume = {696},
       number = {1},
        pages = {385-395},
          doi = {10.1088/0004-637X/696/1/385},
archivePrefix = {arXiv},
       eprint = {0901.4975},
 primaryClass = {astro-ph.GA},
       adsurl = {https://ui.adsabs.harvard.edu/abs/2009ApJ...696..385G}
}

@article{miller15,
   title={CONSTRAINING THE MILKY WAY’S HOT GAS HALO WITH O VII AND O VIII EMISSION LINES},
   volume={800},
   ISSN={1538-4357},
   url={http://dx.doi.org/10.1088/0004-637X/800/1/14},
   DOI={10.1088/0004-637x/800/1/14},
   number={1},
   journal={The Astrophysical Journal},
   publisher={American Astronomical Society},
   author={Miller, Matthew J. and Bregman, Joel N.},
   year={2015},
   month=Feb, pages={14} }

@article{locatelli24,
   title={The warm-hot circumgalactic medium of the Milky Way as seen by eROSITA},
   volume={681},
   ISSN={1432-0746},
   url={http://dx.doi.org/10.1051/0004-6361/202347061},
   DOI={10.1051/0004-6361/202347061},
   journal={Astronomy \& Astrophysics},
   publisher={EDP Sciences},
   author={Locatelli, N. and Ponti, G. and Zheng, X. and Merloni, A. and Becker, W. and Comparat, J. and Dennerl, K. and Freyberg, M. J. and Sasaki, M. and Yeung, M. C. H.},
   year={2024},
   month=Jan, pages={A78} }

@article{fang15,
   title={<i>XMM-NEWTON</i>
                    SURVEY OF LOCAL ${\rm O}\;{\rm VII}$ ABSORPTION LINES IN THE SPECTRA OF ACTIVE GALACTIC NUCLEI},
   volume={217},
   ISSN={1538-4365},
   url={http://dx.doi.org/10.1088/0067-0049/217/2/21},
   DOI={10.1088/0067-0049/217/2/21},
   number={2},
   journal={The Astrophysical Journal Supplement Series},
   publisher={American Astronomical Society},
   author={Fang, Taotao and Buote, David and Bullock, James and Ma, Renyi},
   year={2015},
   month=Apr, pages={21} }

@ARTICLE{Faerman17,
       author = {{Faerman}, Yakov and {Sternberg}, Amiel and {McKee}, Christopher F.},
        title = "{Massive Warm/Hot Galaxy Coronae as Probed by UV/X-Ray Oxygen Absorption and Emission. I. Basic Model}",
      journal = {\apj},
         year = 2017,
        month = jan,
       volume = {835},
       number = {1},
          eid = {52},
        pages = {52},
          doi = {10.3847/1538-4357/835/1/52},
archivePrefix = {arXiv},
       eprint = {1602.00689},
 primaryClass = {astro-ph.GA},
       adsurl = {https://ui.adsabs.harvard.edu/abs/2017ApJ...835...52F}
}

@ARTICLE{yt20,
       author = {{Yamasaki}, Shotaro and {Totani}, Tomonori},
        title = "{The Galactic Halo Contribution to the Dispersion Measure of Extragalactic Fast Radio Bursts}",
      journal = {\apj},
         year = 2020,
        month = jan,
       volume = {888},
       number = {2},
          eid = {105},
        pages = {105},
          doi = {10.3847/1538-4357/ab58c4},
archivePrefix = {arXiv},
       eprint = {1909.00849},
 primaryClass = {astro-ph.HE},
       adsurl = {https://ui.adsabs.harvard.edu/abs/2020ApJ...888..105Y}
}

@ARTICLE{engawa,
       author = {{Lucchini}, Scott and {Abramson}, Cecilia and {Hummels}, Cameron and {Conroy}, Charlie and {Hernquist}, Lars and {Smith}, Aaron},
        title = "{ENhanced Galactic Atmospheres With Arepo: Resolving the CGM at 200 pc with the ENGAWA Simulations}",
      journal = {arXiv e-prints},
         year = 2026,
        month = mar,
          eid = {arXiv:2603.05584},
        pages = {arXiv:2603.05584},
          doi = {10.48550/arXiv.2603.05584},
archivePrefix = {arXiv},
       eprint = {2603.05584},
 primaryClass = {astro-ph.GA},
       adsurl = {https://ui.adsabs.harvard.edu/abs/2026arXiv260305584L}
}

@ARTICLE{grand17,
       author = {{Grand}, Robert J.~J. and {G{\'o}mez}, Facundo A. and {Marinacci}, Federico and {Pakmor}, R{\"u}diger and {Springel}, Volker and {Campbell}, David J.~R. and {Frenk}, Carlos S. and {Jenkins}, Adrian and {White}, Simon D.~M.},
        title = "{The Auriga Project: the properties and formation mechanisms of disc galaxies across cosmic time}",
      journal = {\mnras},
         year = 2017,
        month = may,
       volume = {467},
       number = {1},
        pages = {179-207},
          doi = {10.1093/mnras/stx071},
archivePrefix = {arXiv},
       eprint = {1610.01159},
 primaryClass = {astro-ph.GA},
       adsurl = {https://ui.adsabs.harvard.edu/abs/2017MNRAS.467..179G}
}

@ARTICLE{pillepich18,
       author = {{Pillepich}, Annalisa and {Springel}, Volker and {Nelson}, Dylan and {Genel}, Shy and {Naiman}, Jill and {Pakmor}, R{\"u}diger and {Hernquist}, Lars and {Torrey}, Paul and {Vogelsberger}, Mark and {Weinberger}, Rainer and {Marinacci}, Federico},
        title = "{Simulating galaxy formation with the IllustrisTNG model}",
      journal = {\mnras},
         year = 2018,
        month = jan,
       volume = {473},
       number = {3},
        pages = {4077-4106},
          doi = {10.1093/mnras/stx2656},
archivePrefix = {arXiv},
       eprint = {1703.02970},
 primaryClass = {astro-ph.GA},
       adsurl = {https://ui.adsabs.harvard.edu/abs/2018MNRAS.473.4077P}
}

@ARTICLE{weinberger17,
       author = {{Weinberger}, Rainer and {Springel}, Volker and {Hernquist}, Lars and {Pillepich}, Annalisa and {Marinacci}, Federico and {Pakmor}, R{\"u}diger and {Nelson}, Dylan and {Genel}, Shy and {Vogelsberger}, Mark and {Naiman}, Jill and {Torrey}, Paul},
        title = "{Simulating galaxy formation with black hole driven thermal and kinetic feedback}",
      journal = {\mnras},
         year = 2017,
        month = mar,
       volume = {465},
       number = {3},
        pages = {3291-3308},
          doi = {10.1093/mnras/stw2944},
archivePrefix = {arXiv},
       eprint = {1607.03486},
 primaryClass = {astro-ph.GA},
       adsurl = {https://ui.adsabs.harvard.edu/abs/2017MNRAS.465.3291W}
}

@ARTICLE{nelson19,
       author = {{Nelson}, Dylan and {Springel}, Volker and {Pillepich}, Annalisa and {Rodriguez-Gomez}, Vicente and {Torrey}, Paul and {Genel}, Shy and {Vogelsberger}, Mark and {Pakmor}, Ruediger and {Marinacci}, Federico and {Weinberger}, Rainer and {Kelley}, Luke and {Lovell}, Mark and {Diemer}, Benedikt and {Hernquist}, Lars},
        title = "{The IllustrisTNG simulations: public data release}",
      journal = {Computational Astrophysics and Cosmology},
         year = 2019,
        month = may,
       volume = {6},
       number = {1},
          eid = {2},
        pages = {2},
          doi = {10.1186/s40668-019-0028-x},
archivePrefix = {arXiv},
       eprint = {1812.05609},
 primaryClass = {astro-ph.GA},
       adsurl = {https://ui.adsabs.harvard.edu/abs/2019ComAC...6....2N}
}

@ARTICLE{springel10,
       author = {{Springel}, Volker},
        title = "{E pur si muove: Galilean-invariant cosmological hydrodynamical simulations on a moving mesh}",
      journal = {\mnras},
         year = 2010,
        month = jan,
       volume = {401},
       number = {2},
        pages = {791-851},
          doi = {10.1111/j.1365-2966.2009.15715.x},
archivePrefix = {arXiv},
       eprint = {0901.4107},
 primaryClass = {astro-ph.CO},
       adsurl = {https://ui.adsabs.harvard.edu/abs/2010MNRAS.401..791S}
}

@ARTICLE{arepo,
       author = {{Weinberger}, Rainer and {Springel}, Volker and {Pakmor}, R{\"u}diger},
        title = "{The AREPO Public Code Release}",
      journal = {\apjs},
         year = 2020,
        month = jun,
       volume = {248},
       number = {2},
          eid = {32},
        pages = {32},
          doi = {10.3847/1538-4365/ab908c},
archivePrefix = {arXiv},
       eprint = {1909.04667},
 primaryClass = {astro-ph.IM},
       adsurl = {https://ui.adsabs.harvard.edu/abs/2020ApJS..248...32W}
}

@ARTICLE{colt,
       author = {{Smith}, Aaron and {Safranek-Shrader}, Chalence and {Bromm}, Volker and {Milosavljevi{\'c}}, Milo{\v{s}}},
        title = "{The Lyman {\ensuremath{\alpha}} signature of the first galaxies}",
      journal = {\mnras},
         year = 2015,
        month = jun,
       volume = {449},
       number = {4},
        pages = {4336-4362},
          doi = {10.1093/mnras/stv565},
archivePrefix = {arXiv},
       eprint = {1409.4480},
 primaryClass = {astro-ph.CO},
       adsurl = {https://ui.adsabs.harvard.edu/abs/2015MNRAS.449.4336S}
}

@ARTICLE{vogelsberger20,
       author = {{Vogelsberger}, Mark and {Marinacci}, Federico and {Torrey}, Paul and {Puchwein}, Ewald},
        title = "{Cosmological simulations of galaxy formation}",
      journal = {Nature Reviews Physics},
         year = 2020,
        month = jan,
       volume = {2},
       number = {1},
        pages = {42-66},
          doi = {10.1038/s42254-019-0127-2},
archivePrefix = {arXiv},
       eprint = {1909.07976},
 primaryClass = {astro-ph.GA},
       adsurl = {https://ui.adsabs.harvard.edu/abs/2020NatRP...2...42V}
}

@ARTICLE{suresh19,
       author = {{Suresh}, Joshua and {Nelson}, Dylan and {Genel}, Shy and {Rubin}, Kate H.~R. and {Hernquist}, Lars},
        title = "{Zooming in on accretion - II. Cold circumgalactic gas simulated with a super-Lagrangian refinement scheme}",
      journal = {\mnras},
         year = 2019,
        month = mar,
       volume = {483},
       number = {3},
        pages = {4040-4059},
          doi = {10.1093/mnras/sty3402},
archivePrefix = {arXiv},
       eprint = {1811.01949},
 primaryClass = {astro-ph.GA},
       adsurl = {https://ui.adsabs.harvard.edu/abs/2019MNRAS.483.4040S}
}

@ARTICLE{gible,
       author = {{Ramesh}, Rahul and {Nelson}, Dylan},
        title = "{Zooming in on the circumgalactic medium with GIBLE: Resolving small-scale gas structure in cosmological simulations}",
      journal = {\mnras},
         year = 2024,
        month = feb,
       volume = {528},
       number = {2},
        pages = {3320-3339},
          doi = {10.1093/mnras/stae237},
archivePrefix = {arXiv},
       eprint = {2307.11143},
 primaryClass = {astro-ph.GA},
       adsurl = {https://ui.adsabs.harvard.edu/abs/2024MNRAS.528.3320R}
}

@ARTICLE{tempest,
       author = {{Hummels}, Cameron B. and {Smith}, Britton D. and {Hopkins}, Philip F. and {O'Shea}, Brian W. and {Silvia}, Devin W. and {Werk}, Jessica K. and {Lehner}, Nicolas and {Wise}, John H. and {Collins}, David C. and {Butsky}, Iryna S.},
        title = "{The Impact of Enhanced Halo Resolution on the Simulated Circumgalactic Medium}",
      journal = {\apj},
         year = 2019,
        month = sep,
       volume = {882},
       number = {2},
          eid = {156},
        pages = {156},
          doi = {10.3847/1538-4357/ab378f},
archivePrefix = {arXiv},
       eprint = {1811.12410},
 primaryClass = {astro-ph.GA},
       adsurl = {https://ui.adsabs.harvard.edu/abs/2019ApJ...882..156H}
}

@ARTICLE{foggie,
       author = {{Peeples}, Molly S. and {Corlies}, Lauren and {Tumlinson}, Jason and {O'Shea}, Brian W. and {Lehner}, Nicolas and {O'Meara}, John M. and {Howk}, J. Christopher and {Earl}, Nicholas and {Smith}, Britton D. and {Wise}, John H. and {Hummels}, Cameron B.},
        title = "{Figuring Out Gas \& Galaxies in Enzo (FOGGIE). I. Resolving Simulated Circumgalactic Absorption at 2 {\ensuremath{\leq}} z {\ensuremath{\leq}} 2.5}",
      journal = {\apj},
         year = 2019,
        month = mar,
       volume = {873},
       number = {2},
          eid = {129},
        pages = {129},
          doi = {10.3847/1538-4357/ab0654},
archivePrefix = {arXiv},
       eprint = {1810.06566},
 primaryClass = {astro-ph.GA},
       adsurl = {https://ui.adsabs.harvard.edu/abs/2019ApJ...873..129P}
}

@ARTICLE{surge,
       author = {{van de Voort}, Freeke and {Springel}, Volker and {Mandelker}, Nir and {van den Bosch}, Frank C. and {Pakmor}, R{\"u}diger},
        title = "{Cosmological simulations of the circumgalactic medium with 1 kpc resolution: enhanced H I column densities}",
      journal = {\mnras},
         year = 2019,
        month = jan,
       volume = {482},
       number = {1},
        pages = {L85-L89},
          doi = {10.1093/mnrasl/sly190},
archivePrefix = {arXiv},
       eprint = {1808.04369},
 primaryClass = {astro-ph.GA},
       adsurl = {https://ui.adsabs.harvard.edu/abs/2019MNRAS.482L..85V}
}

@ARTICLE{foggie5,
       author = {{Lochhaas}, Cassandra and {Tumlinson}, Jason and {O'Shea}, Brian W. and {Peeples}, Molly S. and {Smith}, Britton D. and {Werk}, Jessica K. and {Augustin}, Ramona and {Simons}, Raymond C.},
        title = "{Figuring Out Gas \& Galaxies In Enzo (FOGGIE). V. The Virial Temperature Does Not Describe Gas in a Virialized Galaxy Halo}",
      journal = {\apj},
         year = 2021,
        month = dec,
       volume = {922},
       number = {2},
          eid = {121},
        pages = {121},
          doi = {10.3847/1538-4357/ac2496},
archivePrefix = {arXiv},
       eprint = {2102.08393},
 primaryClass = {astro-ph.GA},
       adsurl = {https://ui.adsabs.harvard.edu/abs/2021ApJ...922..121L}
}

@ARTICLE{foggie6,
       author = {{Lochhaas}, Cassandra and {Tumlinson}, Jason and {Peeples}, Molly S. and {O'Shea}, Brian W. and {Werk}, Jessica K. and {Simons}, Raymond C. and {Juno}, James and {Kopenhafer}, Claire and {Augustin}, Ramona and {Wright}, Anna C. and {Acharyya}, Ayan and {Smith}, Britton D.},
        title = "{Figuring Out Gas \& Galaxies in Enzo (FOGGIE). VI. The Circumgalactic Medium of L $^{{\ensuremath{*}}}$ Galaxies Is Supported in an Emergent, Nonhydrostatic Equilibrium}",
      journal = {\apj},
         year = 2023,
        month = may,
       volume = {948},
       number = {1},
          eid = {43},
        pages = {43},
          doi = {10.3847/1538-4357/acbb06},
archivePrefix = {arXiv},
       eprint = {2206.09925},
 primaryClass = {astro-ph.GA},
       adsurl = {https://ui.adsabs.harvard.edu/abs/2023ApJ...948...43L}
}

@ARTICLE{foggie10,
       author = {{Augustin}, Ramona and {Tumlinson}, Jason and {Peeples}, Molly S. and {O'Shea}, Brian W. and {Smith}, Britton D. and {Lochhaas}, Cassandra and {Wright}, Anna C. and {Acharyya}, Ayan and {Werk}, Jessica K. and {Lehner}, Nicolas and {Corlies}, Lauren and {Simons}, Raymond C. and {Howk}, J. Christopher and {O'Meara}, John M.},
        title = "{FOGGIE. X. Characterizing the Small-scale Structure of the Circumgalactic Medium and Its Imprint on Observables}",
      journal = {\apj},
         year = 2025,
        month = nov,
       volume = {993},
       number = {1},
          eid = {52},
        pages = {52},
          doi = {10.3847/1538-4357/ae0462},
archivePrefix = {arXiv},
       eprint = {2501.06551},
 primaryClass = {astro-ph.GA},
       adsurl = {https://ui.adsabs.harvard.edu/abs/2025ApJ...993...52A}
}

@ARTICLE{huang25,
       author = {{Huang}, Yuxin and {Lee}, Khee-Gan and {Libeskind}, Noam I. and {Simha}, Sunil and {Valade}, Aur{\'e}lien and {Prochaska}, J. Xavier},
        title = "{Modelling the cosmic dispersion measure in the D < 120 Mpc Local Universe}",
      journal = {\mnras},
         year = 2025,
        month = apr,
       volume = {538},
       number = {4},
        pages = {2785-2799},
          doi = {10.1093/mnras/staf417},
archivePrefix = {arXiv},
       eprint = {2410.22098},
 primaryClass = {astro-ph.CO},
       adsurl = {https://ui.adsabs.harvard.edu/abs/2025MNRAS.538.2785H}
}

@ARTICLE{wham,
       author = {{Haffner}, L.~M. and {Reynolds}, R.~J. and {Tufte}, S.~L. and {Madsen}, G.~J. and {Jaehnig}, K.~P. and {Percival}, J.~W.},
        title = "{The Wisconsin H{\ensuremath{\alpha}} Mapper Northern Sky Survey}",
      journal = {\apjs},
         year = 2003,
        month = dec,
       volume = {149},
       number = {2},
        pages = {405-422},
          doi = {10.1086/378850},
archivePrefix = {arXiv},
       eprint = {astro-ph/0309117},
 primaryClass = {astro-ph},
       adsurl = {https://ui.adsabs.harvard.edu/abs/2003ApJS..149..405H}
}

@ARTICLE{anderson10,
       author = {{Anderson}, Michael E. and {Bregman}, Joel N.},
        title = "{Do Hot Halos Around Galaxies Contain the Missing Baryons?}",
      journal = {\apj},
         year = 2010,
        month = may,
       volume = {714},
       number = {1},
        pages = {320-331},
          doi = {10.1088/0004-637X/714/1/320},
archivePrefix = {arXiv},
       eprint = {1003.3273},
 primaryClass = {astro-ph.CO},
       adsurl = {https://ui.adsabs.harvard.edu/abs/2010ApJ...714..320A}
}

@ARTICLE{bregman18,
       author = {{Bregman}, Joel N. and {Anderson}, Michael E. and {Miller}, Matthew J. and {Hodges-Kluck}, Edmund and {Dai}, Xinyu and {Li}, Jiang-Tao and {Li}, Yunyang and {Qu}, Zhijie},
        title = "{The Extended Distribution of Baryons around Galaxies}",
      journal = {\apj},
         year = 2018,
        month = jul,
       volume = {862},
       number = {1},
          eid = {3},
        pages = {3},
          doi = {10.3847/1538-4357/aacafe},
archivePrefix = {arXiv},
       eprint = {1803.08963},
 primaryClass = {astro-ph.GA},
       adsurl = {https://ui.adsabs.harvard.edu/abs/2018ApJ...862....3B}
}

@ARTICLE{connor2025,
       author = {{Connor}, Liam and {Ravi}, Vikram and {Sharma}, Kritti and {Ocker}, Stella Koch and {Faber}, Jakob and {Hallinan}, Gregg and {Harnach}, Charlie and {Hellbourg}, Greg and {Hobbs}, Rick and {Hodge}, David and {Hodges}, Mark and {Kosogorov}, Nikita and {Lamb}, James and {Law}, Casey and {Rasmussen}, Paul and {Sherman}, Myles and {Somalwar}, Jean and {Weinreb}, Sander and {Woody}, David and {Konietzka}, Ralf M.},
        title = "{A gas-rich cosmic web revealed by the partitioning of the missing baryons}",
      journal = {Nature Astronomy},
         year = 2025,
        month = jun,
       volume = {9},
        pages = {1226-1239},
          doi = {10.1038/s41550-025-02566-y},
archivePrefix = {arXiv},
       eprint = {2409.16952},
 primaryClass = {astro-ph.CO},
       adsurl = {https://ui.adsabs.harvard.edu/abs/2025NatAs...9.1226C}
}

@ARTICLE{keatingpen,
       author = {{Keating}, Laura C. and {Pen}, Ue-Li},
        title = "{Exploring the dispersion measure of the Milky Way halo}",
      journal = {\mnras},
         year = 2020,
        month = jul,
       volume = {496},
       number = {1},
        pages = {L106-L110},
          doi = {10.1093/mnrasl/slaa095},
archivePrefix = {arXiv},
       eprint = {2001.11105},
 primaryClass = {astro-ph.GA},
       adsurl = {https://ui.adsabs.harvard.edu/abs/2020MNRAS.496L.106K}
}

@ARTICLE{konietzka2025,
       author = {{Konietzka}, Ralf M. and {Connor}, Liam and {Semenov}, Vadim A. and {Beane}, Angus and {Springel}, Volker and {Hernquist}, Lars},
        title = "{Ray-tracing Fast Radio Bursts Through IllustrisTNG: Cosmological Dispersion Measures from Redshift 0 to 5.5}",
      journal = {arXiv e-prints},
         year = 2025,
        month = jul,
          eid = {arXiv:2507.07090},
        pages = {arXiv:2507.07090},
          doi = {10.48550/arXiv.2507.07090},
archivePrefix = {arXiv},
       eprint = {2507.07090},
 primaryClass = {astro-ph.CO},
       adsurl = {https://ui.adsabs.harvard.edu/abs/2025arXiv250707090K}
}

@ARTICLE{mccarty2026,
       author = {{McCarty}, Samuel and {Connor}, Liam and {Konietzka}, Ralf M.},
        title = "{The CGM with local universe FRBs: evidence of strong AGN feedback in a massive elliptical galaxy}",
      journal = {arXiv e-prints},
         year = 2026,
        month = feb,
          eid = {arXiv:2602.16781},
        pages = {arXiv:2602.16781},
          doi = {10.48550/arXiv.2602.16781},
archivePrefix = {arXiv},
       eprint = {2602.16781},
 primaryClass = {astro-ph.GA},
       adsurl = {https://ui.adsabs.harvard.edu/abs/2026arXiv260216781M}
}

@ARTICLE{ne2001,
       author = {{Cordes}, J. M. and {Lazio}, T. J. W.},
        title = "{NE2001.I. A New Model for the Galactic Distribution of Free Electrons and its Fluctuations}",
      journal = {arXiv e-prints},
         year = 2002,
        month = jul,
          eid = {astro-ph/0207156},
        pages = {astro-ph/0207156},
          doi = {10.48550/arXiv.astro-ph/0207156},
archivePrefix = {arXiv},
       eprint = {astro-ph/0207156},
 primaryClass = {astro-ph},
       adsurl = {https://ui.adsabs.harvard.edu/abs/2002astro.ph..7156C}
}

@ARTICLE{ne2025,
       author = {{Ocker}, Stella Koch and {Cordes}, James M.},
        title = "{NE2025: An Updated Electron Density Model for the Galactic Interstellar Medium}",
      journal = {\apj},
         year = 2026,
        month = apr,
       volume = {1002},
       number = {1},
          eid = {3},
        pages = {3},
          doi = {10.3847/1538-4357/ae5825},
archivePrefix = {arXiv},
       eprint = {2602.11838},
 primaryClass = {astro-ph.GA},
       adsurl = {https://ui.adsabs.harvard.edu/abs/2026ApJ..1002....3O}
}

@ARTICLE{orr,
       author = {{Orr}, Matthew E. and {Burkhart}, Blakesley and {Lu}, Wenbin and {Ponnada}, Sam B. and {Hummels}, Cameron B.},
        title = "{Objects May Be Closer than They Appear: Significant Host Galaxy Dispersion Measures of Fast Radio Bursts in Zoom-in Simulations}",
      journal = {\apjl},
         year = 2024,
        month = sep,
       volume = {972},
       number = {2},
          eid = {L26},
        pages = {L26},
          doi = {10.3847/2041-8213/ad725b},
archivePrefix = {arXiv},
       eprint = {2406.03523},
 primaryClass = {astro-ph.GA},
       adsurl = {https://ui.adsabs.harvard.edu/abs/2024ApJ...972L..26O}
}

@ARTICLE{prochaskazhang,
       author = {{Prochaska}, J. Xavier and {Zheng}, Yong},
        title = "{Probing Galactic haloes with fast radio bursts}",
      journal = {\mnras},
         year = 2019,
        month = may,
       volume = {485},
       number = {1},
        pages = {648-665},
          doi = {10.1093/mnras/stz261},
archivePrefix = {arXiv},
       eprint = {1901.11051},
 primaryClass = {astro-ph.GA},
       adsurl = {https://ui.adsabs.harvard.edu/abs/2019MNRAS.485..648P}
}

@ARTICLE{putman21,
       author = {{Putman}, M. E. and {Zheng}, Yong and {Price-Whelan}, Adrian M. and {Grcevich}, Jana and {Johnson}, Amalya C. and {Tollerud}, Erik and {Peek}, J. E. G.},
        title = "{The Gas Content and Stripping of Local Group Dwarf Galaxies}",
      journal = {\apj},
         year = 2021,
        month = may,
       volume = {913},
       number = {1},
          eid = {53},
        pages = {53},
          doi = {10.3847/1538-4357/abe391},
archivePrefix = {arXiv},
       eprint = {2101.07809},
 primaryClass = {astro-ph.GA},
       adsurl = {https://ui.adsabs.harvard.edu/abs/2021ApJ...913...53P}
}

@ARTICLE{ravi,
       author = {{Ravi}, Vikram and {Catha}, Morgan and {Chen}, Ge and {Connor}, Liam and {Cordes}, James M. and {Faber}, Jakob T. and {Lamb}, James W. and {Hallinan}, Gregg and {Harnach}, Charlie and {Hellbourg}, Greg and {Hobbs}, Rick and {Hodge}, David and {Hodges}, Mark and {Law}, Casey and {Rasmussen}, Paul and {Sharma}, Kritti and {Sherman}, Myles B. and {Shi}, Jun and {Simard}, Dana and {Somalwar}, Jean J. and {Squillace}, Reynier and {Weinreb}, Sander and {Woody}, David P. and {Yadlapalli}, Nitika},
        title = "{Deep Synoptic Array Science: A 50 Mpc Fast Radio Burst Constrains the Mass of the Milky Way Circumgalactic Medium}",
      journal = {\aj},
         year = 2025,
        month = jun,
       volume = {169},
       number = {6},
          eid = {330},
        pages = {330},
          doi = {10.3847/1538-3881/adc725},
archivePrefix = {arXiv},
       eprint = {2301.01000},
 primaryClass = {astro-ph.GA},
       adsurl = {https://ui.adsabs.harvard.edu/abs/2025AJ....169..330R}
}

@ARTICLE{ymw15,
       author = {{Yao}, J. M. and {Manchester}, R. N. and {Wang}, N.},
        title = "{A New Electron-density Model for Estimation of Pulsar and FRB Distances}",
      journal = {\apj},
         year = 2017,
        month = jan,
       volume = {835},
       number = {1},
          eid = {29},
        pages = {29},
          doi = {10.3847/1538-4357/835/1/29},
archivePrefix = {arXiv},
       eprint = {1610.09448},
 primaryClass = {astro-ph.HE},
       adsurl = {https://ui.adsabs.harvard.edu/abs/2017ApJ...835...29Y}
}

@article{Das_2020,
   title={Empirical estimates of the Galactic halo contribution to the dispersion measures of extragalactic fast radio bursts using X-ray absorption},
   volume={500},
   ISSN={1365-2966},
   url={http://dx.doi.org/10.1093/mnras/staa3299},
   DOI={10.1093/mnras/staa3299},
   number={1},
   journal={Monthly Notices of the Royal Astronomical Society},
   publisher={Oxford University Press (OUP)},
   author={Das, Sanskriti and Mathur, Smita and Gupta, Anjali and Nicastro, Fabrizio and Krongold, Yair},
   year={2020},
   month=Oct, pages={655–662} }

@article{Cook_2023,
   title={An FRB Sent Me a DM: Constraining the Electron Column of the Milky Way Halo with Fast Radio Burst Dispersion Measures from CHIME/FRB},
   volume={946},
   ISSN={1538-4357},
   url={http://dx.doi.org/10.3847/1538-4357/acbbd0},
   DOI={10.3847/1538-4357/acbbd0},
   number={2},
   journal={The Astrophysical Journal},
   publisher={American Astronomical Society},
   author={Cook, Amanda M. and Bhardwaj, Mohit and Gaensler, B. M. and Scholz, Paul and Eadie, Gwendolyn M. and Hill, Alex S. and Kaspi, Victoria M. and Masui, Kiyoshi W. and Curtin, Alice P. and Dong, Fengqiu Adam and Fonseca, Emmanuel and Herrera-Martin, Antonio and Kaczmarek, Jane and Lanman, Adam E. and Lazda, Mattias and Leung, Calvin and Meyers, Bradley W. and Michilli, Daniele and Pandhi, Ayush and Pearlman, Aaron B. and Pleunis, Ziggy and Ransom, Scott and Rahman, Mubdi and Sand, Ketan R. and Shin, Kaitlyn and Smith, Kendrick and Stairs, Ingrid and Stenning, David C.},
   year={2023},
   month=Mar, pages={58} }

@ARTICLE{donghia16,
       author = {{D'Onghia}, Elena and {Fox}, Andrew J.},
        title = "{The Magellanic Stream: Circumnavigating the Galaxy}",
      journal = {\araa},
         year = 2016,
        month = sep,
       volume = {54},
        pages = {363-400},
          doi = {10.1146/annurev-astro-081915-023251},
archivePrefix = {arXiv},
       eprint = {1511.05853},
 primaryClass = {astro-ph.GA},
       adsurl = {https://ui.adsabs.harvard.edu/abs/2016ARA&A..54..363D}
}

@ARTICLE{lucchini20,
       author = {{Lucchini}, S. and {D'Onghia}, E. and {Fox}, A.~J. and {Bustard}, C. and {Bland-Hawthorn}, J. and {Zweibel}, E.},
        title = "{The Magellanic Corona as the key to the formation of the Magellanic Stream.}",
      journal = {\nat},
         year = 2020,
        month = jan,
       volume = {585},
        pages = {203-206},
          doi = {10.1038/s41586-020-2663-4},
archivePrefix = {arXiv},
       eprint = {2009.04368},
 primaryClass = {astro-ph.GA},
       adsurl = {https://ui.adsabs.harvard.edu/abs/2020Natur.585..203L}
}

@ARTICLE{dk22,
       author = {{Krishnarao}, Dhanesh and {Fox}, Andrew J. and {D'Onghia}, Elena and {Wakker}, Bart P. and {Cashman}, Frances H. and {Howk}, J. Christopher and {Lucchini}, Scott and {French}, David M. and {Lehner}, Nicolas},
        title = "{Observations of a Magellanic Corona}",
      journal = {\nat},
         year = 2022,
        month = sep,
       volume = {609},
       number = {7929},
        pages = {915-918},
          doi = {10.1038/s41586-022-05090-5},
archivePrefix = {arXiv},
       eprint = {2209.15017},
 primaryClass = {astro-ph.GA},
       adsurl = {https://ui.adsabs.harvard.edu/abs/2022Natur.609..915K}
}

@Article{hunter07,
  Author    = {Hunter, J. D.},
  Title     = {Matplotlib: A 2D graphics environment},
  Journal   = {Computing in Science \& Engineering},
  Volume    = {9},
  Number    = {3},
  Pages     = {90--95},
  publisher = {IEEE COMPUTER SOC},
  doi       = {10.1109/MCSE.2007.55},
  year      = 2007
}

@Article{         harris20,
 title         = {Array programming with {NumPy}},
 author        = {Charles R. Harris and K. Jarrod Millman and St{\'{e}}fan J.
                 van der Walt and Ralf Gommers and Pauli Virtanen and David
                 Cournapeau and Eric Wieser and Julian Taylor and Sebastian
                 Berg and Nathaniel J. Smith and Robert Kern and Matti Picus
                 and Stephan Hoyer and Marten H. van Kerkwijk and Matthew
                 Brett and Allan Haldane and Jaime Fern{\'{a}}ndez del
                 R{\'{i}}o and Mark Wiebe and Pearu Peterson and Pierre
                 G{\'{e}}rard-Marchant and Kevin Sheppard and Tyler Reddy and
                 Warren Weckesser and Hameer Abbasi and Christoph Gohlke and
                 Travis E. Oliphant},
 year          = {2020},
 month         = sep,
 journal       = {Nature},
 volume        = {585},
 number        = {7825},
 pages         = {357--362},
 doi           = {10.1038/s41586-020-2649-2},
 publisher     = {Springer Science and Business Media {LLC}},
 url           = {https://doi.org/10.1038/s41586-020-2649-2}
}

@ARTICLE{hummels17,
    author = {{Hummels}, C.~B. and {Smith}, B.~D. and {Silvia}, D.~W.},
    title = "{Trident: A Universal Tool for Generating Synthetic Absorption Spectra from Astrophysical Simulations}",
    journal = {\apj},
    archivePrefix = "arXiv",
    eprint = {1612.03935},
    primaryClass = "astro-ph.IM",
    year = 2017,
    month = sep,
    volume = 847,
    eid = {59},
    pages = {59},
    doi = {10.3847/1538-4357/aa7e2d},
    adsurl = {http://adsabs.harvard.edu/abs/2017ApJ...847...59H}
}

@ARTICLE{turk11,
       author = {{Turk}, Matthew J. and {Smith}, Britton D. and {Oishi}, Jeffrey S. and {Skory}, Stephen and {Skillman}, Samuel W. and {Abel}, Tom and {Norman}, Michael L.},
        title = "{yt: A Multi-code Analysis Toolkit for Astrophysical Simulation Data}",
      journal = {\apjs},
         year = 2011,
        month = jan,
       volume = {192},
       number = {1},
          eid = {9},
        pages = {9},
          doi = {10.1088/0067-0049/192/1/9},
archivePrefix = {arXiv},
       eprint = {1011.3514},
 primaryClass = {astro-ph.IM},
       adsurl = {https://ui.adsabs.harvard.edu/abs/2011ApJS..192....9T}
}

@misc{hoffmann2026ihaloconstrainingmilky,
      title={I can see your halo: Constraining the Milky Way halo DM with FRB population studies}, 
      author={Jordan Hoffmann and Clancy James and Jason Xavier Prochaska and Marcin Glowacki},
      year={2026},
      eprint={2601.05496},
      archivePrefix={arXiv},
      primaryClass={astro-ph.GA},
      url={https://arxiv.org/abs/2601.05496}, 
}

@misc{liu2026investigatinganisotropydispersionmeasure,
      title={Investigating the Anisotropy of Dispersion Measure Contribution from the Galactic Halo by Using Fast Radio Bursts}, 
      author={Yang Liu and Bao Wang and Puxun Wu and Jun-Jie Wei and Xue-Feng Wu},
      year={2026},
      eprint={2601.02849},
      archivePrefix={arXiv},
      primaryClass={astro-ph.GA},
      url={https://arxiv.org/abs/2601.02849}, 
}

@ARTICLE{salem15,
       author = {{Salem}, Munier and {Besla}, Gurtina and {Bryan}, Greg and {Putman}, Mary and {van der Marel}, Roeland P. and {Tonnesen}, Stephanie},
        title = "{Ram Pressure Stripping of the Large Magellanic Cloud's Disk as a Probe of the Milky Way's Circumgalactic Medium}",
      journal = {\apj},
         year = 2015,
        month = dec,
       volume = {815},
       number = {1},
          eid = {77},
        pages = {77},
          doi = {10.1088/0004-637X/815/1/77},
archivePrefix = {arXiv},
       eprint = {1507.07935},
 primaryClass = {astro-ph.GA},
       adsurl = {https://ui.adsabs.harvard.edu/abs/2015ApJ...815...77S}
}

@article{healpy,
  doi = {10.21105/joss.01298},
  url = {https://doi.org/10.21105/joss.01298},
  year = {2019},
  month = mar,
  publisher = {The Open Journal},
  volume = {4},
  number = {35},
  pages = {1298},
  author = {Andrea Zonca and Leo Singer and Daniel Lenz and Martin Reinecke and Cyrille Rosset and Eric Hivon and Krzysztof Gorski},
  title = {healpy: equal area pixelization and spherical harmonics transforms for data on the sphere in Python},
  journal = {Journal of Open Source Software}
}

@ARTICLE{healpix,
   author = {{G{\'o}rski}, K.~M. and {Hivon}, E. and {Banday}, A.~J. and 
	{Wandelt}, B.~D. and {Hansen}, F.~K. and {Reinecke}, M. and 
	{Bartelmann}, M.},
    title = "{HEALPix: A Framework for High-Resolution Discretization and Fast Analysis of Data Distributed on the Sphere}",
  journal = {\apj},
   eprint = {arXiv:astro-ph/0409513},
     year = 2005,
    month = apr,
   volume = 622,
    pages = {759-771},
      doi = {10.1086/427976},
   adsurl = {http://adsabs.harvard.edu/abs/2005ApJ...622..759G}
}
	\bibliographystyle{aasjournal7}
	
\end{document}